\documentclass[traditabstract]{aa} 

\usepackage{graphicx}
\usepackage{txfonts}
\usepackage{longtable}
\begin{document}

   \title{A Virgo Environmental Survey Tracing Ionised Gas Emission (VESTIGE).V. Properties of the ionised gas filament of M87\thanks{Based on observations obtained with
   MegaPrime/MegaCam, a joint project of CFHT and CEA/DAPNIA, at the Canada-French-Hawaii Telescope
   (CFHT) which is operated by the National Research Council (NRC) of Canada, the Institut National
   des Sciences de l'Univers of the Centre National de la Recherche Scientifique (CNRS) of France and
   the University of Hawaii. Based on observations made with ESO Telescopes at the La Silla Paranal Observatory under programme ID 60.A-9312.}
      }
   \subtitle{}
  \author{A. Boselli\inst{1},  
          M. Fossati\inst{2,3},
	  A. Longobardi\inst{4},
	  G. Consolandi\inst{5},
	  P. Amram\inst{1},
	  M. Sun\inst{6},
	  P. Andreani\inst{7},  
	  M. Boquien\inst{8},
	  J. Braine\inst{9},
	  F. Combes\inst{10,11},
          P. C{\^o}t{\'e}\inst{12},
          J.C. Cuillandre\inst{13},
	  P.A. Duc\inst{14},
	  E. Emsellem\inst{7},
          L. Ferrarese\inst{12},
	  G. Gavazzi\inst{15},
          S. Gwyn\inst{11}, 
	  G. Hensler\inst{16},
	  E.W. Peng\inst{17,4},
	  H. Plana\inst{18},
	  J. Roediger\inst{12},
	  R. Sanchez-Janssen\inst{19},
	  M. Sarzi\inst{20},
	  P. Serra\inst{21},
	  G. Trinchieri\inst{5}  
       }

\institute{     
                Aix Marseille Univ, CNRS, CNES, LAM, Marseille, France
                \email{alessandro.boselli@lam.fr}
        \and
                Max-Planck-Institut f\"{u}r Extraterrestrische Physik, Giessenbachstrasse, 85748, Garching, Germany 
	\and  
                Institute for Computational Cosmology and Centre for Extragalactic Astronomy, Department of Physics, Durham University, South Road, Durham DH1 3LE, UK
		\email{matteo.fossati@durham.ac.uk}
        \and
		Kavli Institute for Astronomy and Astrophysics, Peking University, Beijing 100871, PR China
		\email{alongobardi@pku.edu.cn}
	\and
		INAF - Osservatorio Astronomico di Brera, via Brera 28, 20159 Milano, Italy
        \and
		Department of Physics and Astronomy, University of Alabama in Huntsville, Huntsville, AL 35899, USA
 	\and
 		European Southern Observatory, Karl-Schwarzschild-Strasse 2, 85748, Garching, Germany 
	\and
		Centro de Astronom\'a (CITEVA), Universidad de Antofagasta, Avenida Angamos 601, Antofagasta, Chile
	\and
		Laboratoire d'Astrophysique de Bordeaux, Univ. Bordeaux, CNRS, B18N, all\'ee Geoffroy Saint-Hilaire, 33615 Pessac, France
	\and	
		College de France, 11 Pl. M. Berthelot, F-75005 Paris, France 
        \and
                LERMA, Observatoire de Paris, CNRS, PSL Research University, Sorbonne Universit\'es, UPMC Univ. Paris 06, F-75014 Paris, France
        \and  
                NRC Herzberg Astronomy and Astrophysics, 5071 West Saanich Road, Victoria, BC, V9E 2E7, Canada
        \and        
		AIM, CEA, CNRS, Universit\' Paris-Saclay, Universit\'e Paris Diderot, Sorbonne Paris Cit\'e, Observatoire de Paris, PSL University, F-91191 Gif-sur-Yvette Cedex, France
	\and
		Observatoire Astronomique de Strasbourg, UMR 7750, 11, rue de l'Universit\'e, 67000, Strasbourg, France
	\and
		Universit\'a di Milano-Bicocca, piazza della scienza 3, 20100 Milano, Italy
	\and
		Department of Astrophysics, University of Vienna, T\"urkenschanzstrasse 17, 1180 Vienna, Austria
	\and
		Department of Astronomy, Peking University, Beijing 100871, PR China
	\and
		Laborat\'orio de Astrof\'isica Te\'orica e Observacional, Universidade Estadual de Santa Cruz - 45650-000, Ilh\'eus-BA, Brasil
	\and
		UK Astronomy Technology Centre, Royal Observatory Edinburgh, Blackford Hill, Edinburgh, EH9 3HJ, UK
	\and
		Centre for Astrophysics Research, University of Hertfordshire, Hatfield AL10 9AB, UK
	\and
		Osservatorio Astronomico di Cagliari, via della scienza 5, 09047 Selargius, Cagliari, Italy
                 }

\authorrunning{Boselli et al.}
\titlerunning{VESTIGE}

   \date{}

 
  \abstract  
{We have observed the giant elliptical galaxy M87 during the Virgo Environmental Survey Tracing Galaxy Evolution (VESTIGE), 
a blind narrow-band H$\alpha$+[NII] imaging survey of the Virgo cluster carried out with MegaCam at the Canada French Hawaii Telescope (CFHT).
The deep narrow-band image confirmed the presence of a filament of ionised gas extending up to $\simeq$ 
3 kpc in the north-western direction and $\simeq$ 8 kpc to the south-east, with a couple of plumes of ionised gas, the weakest of which, 
at $\simeq$ 18 kpc from the nucleus, was previously unknown. The analysis of deep optical images taken from the NGVS survey confirms that
this gas filament is associated with dust seen in absorption which is now detected up to $\simeq$ 2.4 kpc from the nucleus.
We have also analysed the physical and kinematical properties of the ionised gas filament using deep IFU MUSE data covering the central 4.8 $\times$
4.8 kpc$^2$ of the galaxy. 
The spectroscopic data confirms a perturbed kinematics of the ionised gas, with differences in velocity of $\simeq$ 700-800 km s$^{-1}$ on scales of $\lesssim$ 1 kpc.
The analysis of 2D diagnostic diagrams and the observed relationship between the shock-sensitive [OI]/H$\alpha$ line ratio and the velocity dispersion of the gas
suggest that the gas is shock-ionised. 
 }
   {}
   {}
   {}
   {}
   {}


   \maketitle
%

\section{Introduction}

Giant ellipticals in the centre of rich clusters are among the most massive and luminous objects in the local Universe.
They share with the cluster several properties such as a diffuse X-ray and stellar emission and a similar rest-frame velocity (e.g. Sarazin 1986). 
They are often used to trace the properties of the cluster dark matter halo in the study of the galaxy-halo connection (galaxy conformity),
making them unique systems among all galaxy populations.
Located in the deep of the potential well of a large dynamic structure, they have been formed through the accretion of 
baryonic matter in its different forms, from the diffuse intracluster medium through cooling flows (Cowie \& Binney 1977, Fabian \& Nulsen 1977, 
Fabian 1994), to the cannibalism of other galaxies orbiting within the 
cluster (Ostriker \& Tremaine 1975. White 1976, Malumuth \& Richstone 1984, Merritt 1985, Byrd \& Valtonen 1990), and through multiple merging events that
occurred during the formation of the cluster itself (e.g. De Lucia \& Blaizot 2007, Boselli et al. 2014a).

Among these objects, the elliptical galaxy M87 (NGC 4486, 3C274, Virgo A) in the centre of Virgo is by far the most studied massive galaxy 
in the centre of a cluster, but also one of the most studied galaxies ever. 
At a distance of only 16.5 Mpc (Mei et al. 2007), M87 can be fully resolved at almost all frequencies, from the X-rays to the radio centimetric. 
At this distance, one arcsec corresponds to $\simeq$ 80 pc, a small size compared to the optical extension of the galaxy which has an isophotal radius of
$\simeq$ 25 kpc (Cortese et al. 2012).
This galaxy is a powerful radio source (Baade \& Minkowski 1954) characterised by a prominent jet and two extended radio lobes (e.g. Hines et al. 1989, Owen et al. 2000). 
It also has a characteristic X-ray emission showing long filaments of hot gas extending up to 22 kpc to the east and 28 kpc to the south embedded in a diffuse 
halo (e.g. Young et al. 2002, Forman et al. 2007, Churazov et al. 2008, Werner et al. 2010). 

Similar filamentary structures have been observed also in narrow-band H$\alpha$ imaging (Arp 1967, Ford \& Butcher 1979, Baum et al. 1988, 
Heckman et al. 1989, Sparks et al. 1993, Gavazzi et al. 2000, Werner et al. 2010), in different UV lines with HST (Sparks et al. 2009, Anderson \& Sunayaev
2018) and in the [CII]$\lambda$158 $\mu$m line with \textit{Herschel} (Werner et al. 2013), indicating the multitemperature nature of the gas.
In particular, the narrow-band H$\alpha$ imaging data revealed a very complex filament extending to the south-east up to $\simeq$ 8 kpc 
from the nucleus, with a few patchy regions, whose nature and origin are still not fully understood.

The H$\alpha$ emission line traces the distribution of the ionised hydrogen with a typical temperature of $\simeq$ 10$^4$ K
(Osterbrock \& Ferland 2006). For this reason this line has been often used to look for cooling flows in the centre of rich clusters
of galaxies (e.g. Conselice et al. 2001). In spiral galaxies, where the gas is ionised by the young stellar populations, 
the H$\alpha$ line is an excellent tracer of the recent star formation activity (Kennicutt 1998, Boselli et al. 2009). 
In early-type galaxies the gas can be ionised by a residual star formation activity (Kaviraj et al. 2007, Gavazzi et al. 2018) or 
by a low level ionisation due to hot, evolved (post-asymptotic giant branch) stars (Gomes et al. 2016, Belfiore et al. 2016), the same stellar population
responsible for the UV-upturn observed in the far ultraviolet (O'Connell 1999, Boselli et al. 2005). In M87 the gas can be 
also shock-excited by the central AGN or by the radio jet and counter-jet in the expanding lobes 
(Dopita \& Sutherland 1995, Dopita et al. 1997), or heated by thermal conduction (Sparks et al. 2004; McDonald et al. 2010). 
The gas filament could be hot gas accreted from the intracluster medium cooling to $T$ $\simeq$ 10$^4$ K
or cold gas stripped from a gas-rich cluster galaxy after a gravitational interaction or lost 
during a recent ram pressure stripping event while crossing the halo of M87 (Sparks et al. 1993, Weil et al. 1997, Mayer et al. 2006). 

We have recently observed M87 as part of VESTIGE (A Virgo Environmental Survey Tracing Ionised Gas Emission), a CFHT large programme designed to 
make a blind deep narrow-band H$\alpha$ imaging survey of the whole Virgo cluster up to its virial radius (104 deg$^2$, Boselli et al. 2018a). 
The core of the cluster, including M87 and its surrounding regions, has been mapped during the 2017A observing campaign. Thanks to
a tuned observing strategy and a specific data reduction technique optimised to detect extended, low surface brightness features in the continuum-subtracted
narrow-band images, the VESTIGE data 
overcome in terms of sensitivity and angular resolution those available in the literature, passing from a limiting surface brightness of
$\Sigma(H\alpha+[NII])$ $\simeq$ 10$^{-16}$ erg s$^{-1}$ cm$^{-2}$ arcsec$^{-2}$ to $\Sigma(H\alpha+[NII])$ $\simeq$ 
5 $\times$ 10$^{-18}$ erg s$^{-1}$ cm$^{-2}$ arcsec$^{-2}$. The success of this original observing and data reduction strategy has been 
proven by the detection of extended tails of ionised gas in several Virgo spirals (Boselli et al. 2016, 2018a,b, Fossati et al. 2018).
In this paper we present these new narrow-band imaging data obtained during the VESTIGE survey. 
We also present new photometric and spectroscopic data obtained at other frequencies useful for the study of the ionised gas filament. These include 
a deep spectroscopic IFU MUSE field in the central 1\arcmin $\times$ 1\arcmin of the galaxy taken as part of the first instrument verification run (Emsellem et al.
2014), GALEX UV from the GUViCS survey (Boselli et al. 2011), and optical images from the NGVS survey (Ferrarese et al. 2012).
We then compare this new dataset with data at other frequencies available in the literature 
to have a complete view of the different phases of the interstellar medium (ISM) along the filament and in the 
surrounding intracluster medium (ICM). Since the properties of the multitemperature gas of M87 have been already studied in detail 
in the past, we refer the reader to these publications (Sparks et al. 1993, 2004; Churazov et al. 2001, 2008; Young et al. 2002; 
Forman et al. 2007, 2017; Werner et al. 2013; Simionescu et al. 2018).
Here we limit our analysis to the novelties brought by our new dataset.  
The paper is structured as follows: the VESTIGE and MUSE observations and data reduction are presented in Sect. 2, along with the multifrequency data 
available in the literature. In Sect. 3 we analyse the new dataset through a comparative multifrequency analysis. 
Discussion and conclusions are given in Sects. 4 and 5. Consistently with the other VESTIGE publications, we assume a distance of 16.5 Mpc 
for M87 and the Virgo cluster (Gavazzi et al. 1999, Mei et al. 2007).

\section{Observations and data reduction}

\subsection{Narrow-band imaging}

M87 and its surrounding regions have been mapped during the first observing semester of the VESTIGE project (2017A;
Boselli et al. 2018a for details). The observations have been taken using MegaCam at the CFHT through the narrow-band filter MP9603
($\lambda_c$ = 6591 \AA; $\Delta \lambda$ = 106 \AA) which includes at the redshift of M87 ($vel$ = 1292 km s$^{-1}$, GOLDMine) 
the H$\alpha$ line ($\lambda$ = 6563 \AA) and the two [NII] lines
($\lambda$ =6548, 6583 \AA), 
and through the $r$-band filter for the stellar continuum subtraction. 
MegaCam is composed of 40 2048$\times$4096 pixels CCDs with a pixel size of 0.187$^{\prime\prime}$ on the sky.
The galaxy has been covered with 12 independent frames with a large dithering selected for the whole VESTIGE blind survey (15 arcmin in
R.A., 20 arcmin in dec.). Each single exposure was of 600 sec in the narrow-band H$\alpha$ filter and 60 sec in the broad-band $r$ filter.
The observations have been carried out in good seeing conditions, with a typical seeing of 0.90 and 0.83 arcsec in the 
narrow-band and broad-band stacked frames, respectively.

The data have been reduced following the VESTIGE standard procedures presented in Boselli et al. (2018a) 
and Fossati et al. (in prep.). This consists in using Elixir-LSB (Ferrarese et al. 2012), a data reduction pipeline
expressly designed to detect extended low surface brightness features associated with perturbed galaxies through an accurate 
determination and subtraction of any extended feature in the sky background.

The photometric calibration of the $r$-band frames has been derived following the standard MegaCam calibration procedures (Gwyn 2008), as
extensively described in Boselli et al. (2018a). The calibration in the narrow-band filter has been done as described in Fossati et al. (in
prep.). The typical uncertainty of VESTIGE in both bands is of $\lesssim$ 0.02-0.03 mag. 

The subtraction of the stellar continuum for the determination of the image of the emitting gas
is critical in early-type galaxies such as M87 which are characterised by a strong stellar 
emission in their centre, where the ionised gas is expected to be detected. 

Given the wide band of the $r$ filter
($\lambda_c$ = 6404 \AA; $\Delta \lambda$ = 1480 \AA) and the slight difference in the peak transmissivity of the two bands, 
the derivation of the stellar continuum in the narrow-band from the $r$-band depends on the spectral properties of the 
emitting source (Spector et al. 2012), thus on the colour that in early-type galaxies is known to change radially from the core 
to the periphery (Boroson et al. 1983, Franx \& Illingworth 1990, Suh et al. 2010, Roediger et al. 2011). As extensively described in Fossati et al. (in prep.),
we have used several hundred thousands of unsaturated stars 
detected in the H$\alpha$ narrow-band and in the $r$- and $g$-bands in the science frames to calibrate an empirical relation between the colour of the stars (expressed in AB magnitudes) and 
the normalisation factor:

\begin{equation}
{\frac{r}{H\alpha+[NII]} = r - 0.1713 \times (g-r) + 0.0717}
\end{equation}

\begin{figure*}
   \centering
   \includegraphics[width=1.04\textwidth]{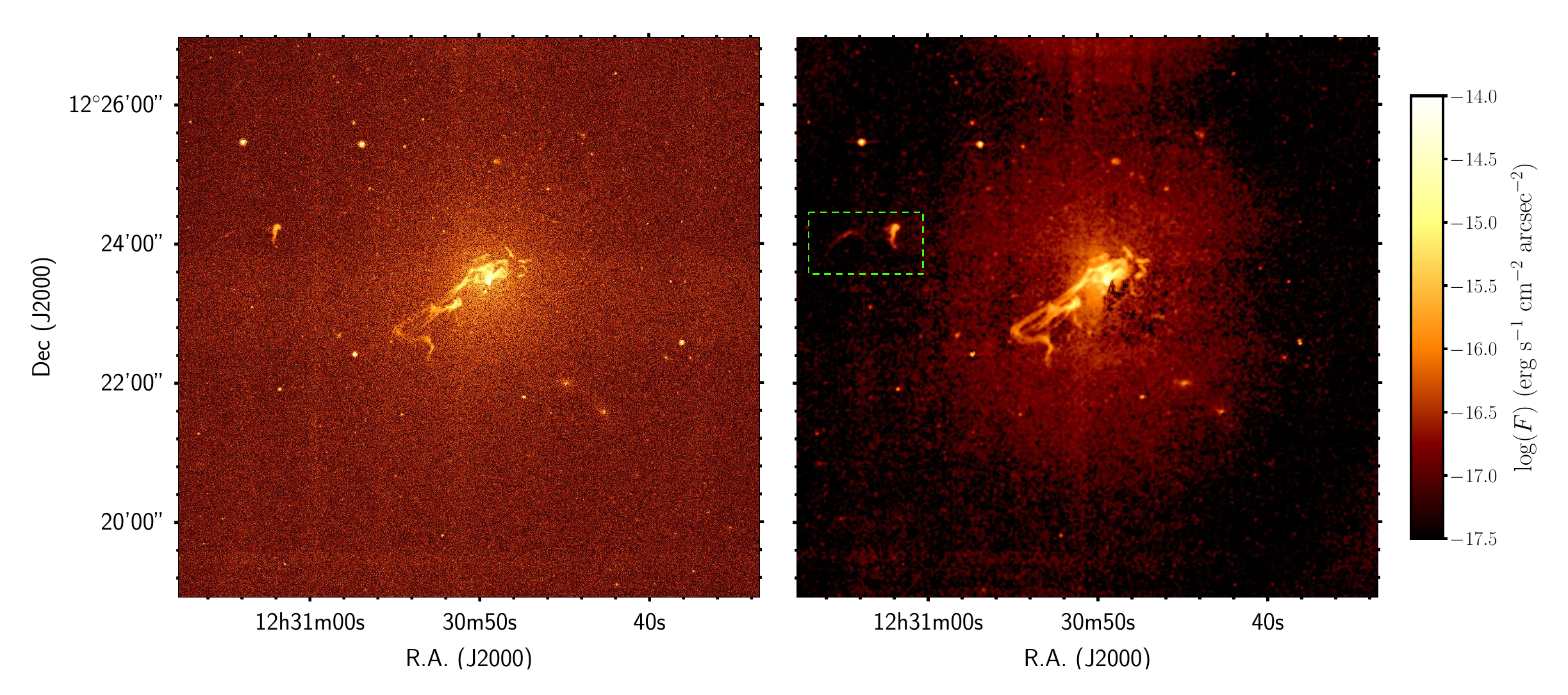}
   \caption{Left: continuum-subtracted H$\alpha$+[NII] image of M87 at full resolution. Right: continuum-subtracted H$\alpha$+[NII] image of M87 
   after a 15$\times$15 boxcar smoothing, corresponding to $\simeq$ 2.8 arcsec resolution. At the distance of M87 (16.5 Mpc), 1 arcsec = 80 pc.
   The green dashed box indicates the region zoomed in Fig. \ref{M87plume}. Smoothing reduces the noise in the outer parts, where the surface brightness sensitivity reaches 
   $\Sigma(H\alpha+[NII])$ $\simeq$ 5 $\times$ 10$^{-18}$ erg s$^{-1}$ cm$^{-2}$ arcsec$^{-2}$.
 }
   \label{M87Hasmo}%
   \end{figure*}

We applied this normalisation pixel by pixel on the stacked frame before the subtraction of the stellar continuum. 
The $g-r$ colour map of the galaxy has been derived using the $g$-band frame taken with MegaCam during the NGVS survey (Ferrarese et al.
2012). To avoid the introduction of any extra noise in the sky regions, where there is no stellar continuum, this colour-dependent normalisation is applied
only whenever the signal in the $r$- and $g$-bands has a signal-to-noise $S/N$ $>$ 1. The continuum-subtracted image is then multiplied by the filter
width (106 \AA) to obtain the values of the line flux per pixel. The resulting H$\alpha$+[NII] stellar continuum 
subtracted image of M87 is shown in Fig. \ref{M87Hasmo}. The quality of this image is excellent: indeed, there are no extended dark regions 
often observed whenever the stellar continuum is over subtracted. 

The total observed flux of the galaxy (corrected for Galactic attenuation using the Schlegel et al. (1998) dust map recalibrated with Schlafly \& Finkbeiner (2011) and 
assuming the Fitzpatrick et al. (1999) Galactic attenuation curve) is log $f(H\alpha+[NII])$ = -11.66 $\pm$ 0.04 erg s$^{-1}$ cm$^{-2}$ and the equivalent width 
H$\alpha$+[NII]E.W. = 2.1 $\pm$ 0.2 \AA, and is dominated by the emission of the filament.

\subsection{MUSE spectroscopy}

\begin{figure*}
   \centering
   \includegraphics[width=1\textwidth]{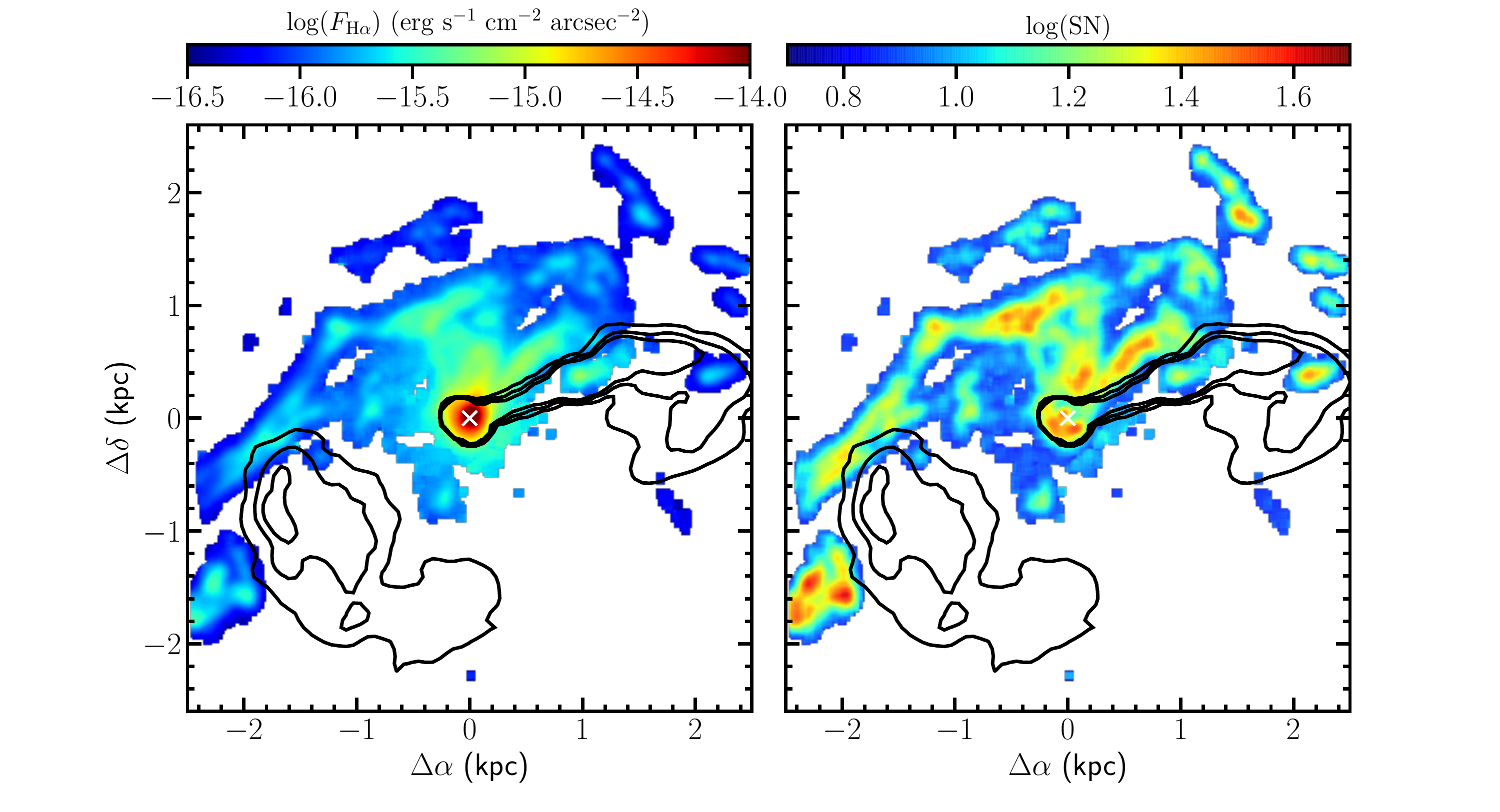}
   \caption{The H$\alpha$ emission line map (left) and its signal-to-noise (right) of the ionised gas derived from MUSE for a $S/N$ $>$5. 
   The white cross in the centre of the frame indicates the position of the nucleus. The contour levels indicate the jet and counter-jet 
   determined from 115 GHz ALMA data.
 }
   \label{MUSE}%
   \end{figure*}

M87 has been observed in 2014 with MUSE during the first instrument science verification run at the VLT. Details of the observations and data reduction are 
given in Emsellem et al. (2014) and Sarzi et al. (2018). MUSE provides a spectroscopic data cube on a rectangular 1\arcmin $\times$1\arcmin field with 
spaxels of 0.2\arcsec $\times$0.2\arcsec. The spectra cover the 4800-9000 \AA ~ range, with an instrumental resolution of $\sim$ 60 km s$^{-1}$ at 5500 \AA.
Two 1800 sec exposures have been taken on the centre of the galaxy. Each exposure was followed by a 900 sec exposure in the outskirts of the galaxy for an accurate
determination of the sky background. The data have been reduced using the v1.6 MUSE standard pipeline, as described in Sarzi et al. (2018).
To subtract the stellar continuum emission and remove any possible contamination due to the underlying stellar absorption on the main 
Balmer lines we fitted the spectra using pPXF and GANDALF (Cappellari \& Emsellem 2004; Sarzi et al. 2006). This code simultaneously models 
the stellar continuum and the emission lines in individual spaxels. The stellar continuum is modelled with a superposition of stellar templates
convolved by the stellar line-of-sight velocity distribution, while the gas emission and kinematics are derived assuming a Gaussian
line profile. Stellar templates are constructed using the MILES library (Vazdekis et al. 2010).
We then fit the emission line (H$\beta$$\lambda$4861, [OIII]$\lambda$4959,5007,[OI]$\lambda$6300,6364,
[NII]$\lambda$6548,6583, H$\alpha$$\lambda$6563, [SII]$\lambda$6716,6731, [SIII]$\lambda$9069) using the KUBEVIZ code as in Fossati et al. (2016). 
Groups of lines are fitted simultaneously using a combination of 
1D Gaussian functions with fixed relative velocities. The noise is measured from the `stat' data cube and renormalised on the line fluxes to take into account
the correlated noise introduced by resampling and smoothing, as extensively described in Fossati et al. (2016). 
Figure \ref{MUSE} shows the H$\alpha$ emission line map and its signal-to-noise map derived for a $S/N$$>$5.  

\subsection{UV}

UV data can be used to identify the nature of the ionising source of the gas filaments observed in Fig. \ref{M87Hasmo}.
Several GALEX exposures, available in both the $FUV$ ($\lambda_c$ = 1539 \AA) and $NUV$ ($\lambda_c$ = 2316 \AA) bands, have been collected during the GUViCS survey
of the cluster (Boselli et al. 2011). We have combined the available frames with a sufficient signal to noise and obtained deep UV images of M87
with a typical integration time of 3278 sec in the $FUV$ and 16267 sec in the $NUV$, and used them to create the $FUV$/$NUV$ colour image. 
M87 has a diffuse light distribution in the UV, witnessing the presence of an evolved stellar population. The only evident structure in the image is the nucleus and the jet, which is
present in both the $FUV$ and $NUV$ bands. The UV colour image does not show any clumpy structure which can be associated with a young stellar population. It is thus likely 
that, as in other massive early-type galaxies, the UV emission of M87 is dominated by evolved stellar populations (UV upturn; O'Connell 1999; Boselli et al. 2005, 2014a).

\subsection{Visible}

Optical data are essential to see the distribution of dust seen in absorption. For M87 this is critical since in this object
dust outside the nuclear region has never been detected in emission even in the most recent 
\textit{Spitzer} (Perlman et al. 2007, Bendo et al. 2012, Ciesla et al. 2014) and \textit{Herschel} (Ciesla et al. 2012, Cortese et al. 2014) 
observations\footnote{The detection reported in these works is due to the synchrotron emission associated with the radio galaxy and its jet (Baes et al. 2010, Boselli et al. 2010).}. 
M87 has been observed with HST during the ACS Virgo Cluster Survey (VCS) in the F475W and F850LP bands ($\sim$ Sloan $g$ and $z$) (C\^ot\'e et al. 2004)
and in the F275W, F606W, and F814W bands by Bellini et al. (2015). 
The $g-z$ colour image of the central $\sim$ 1\arcmin$\times$1\arcmin given in Ferrarese et al. (2006, their Fig. 1) and the galaxy-subtracted F275W, F606W, F814W colour image
of Bellini et al. (2015, their Fig. 15) show a very faint filament of dust  
located in the western direction close to the position of the jet and patches of dust close to the centre.

\begin{figure}
   \centering
   \includegraphics[width=0.5\textwidth]{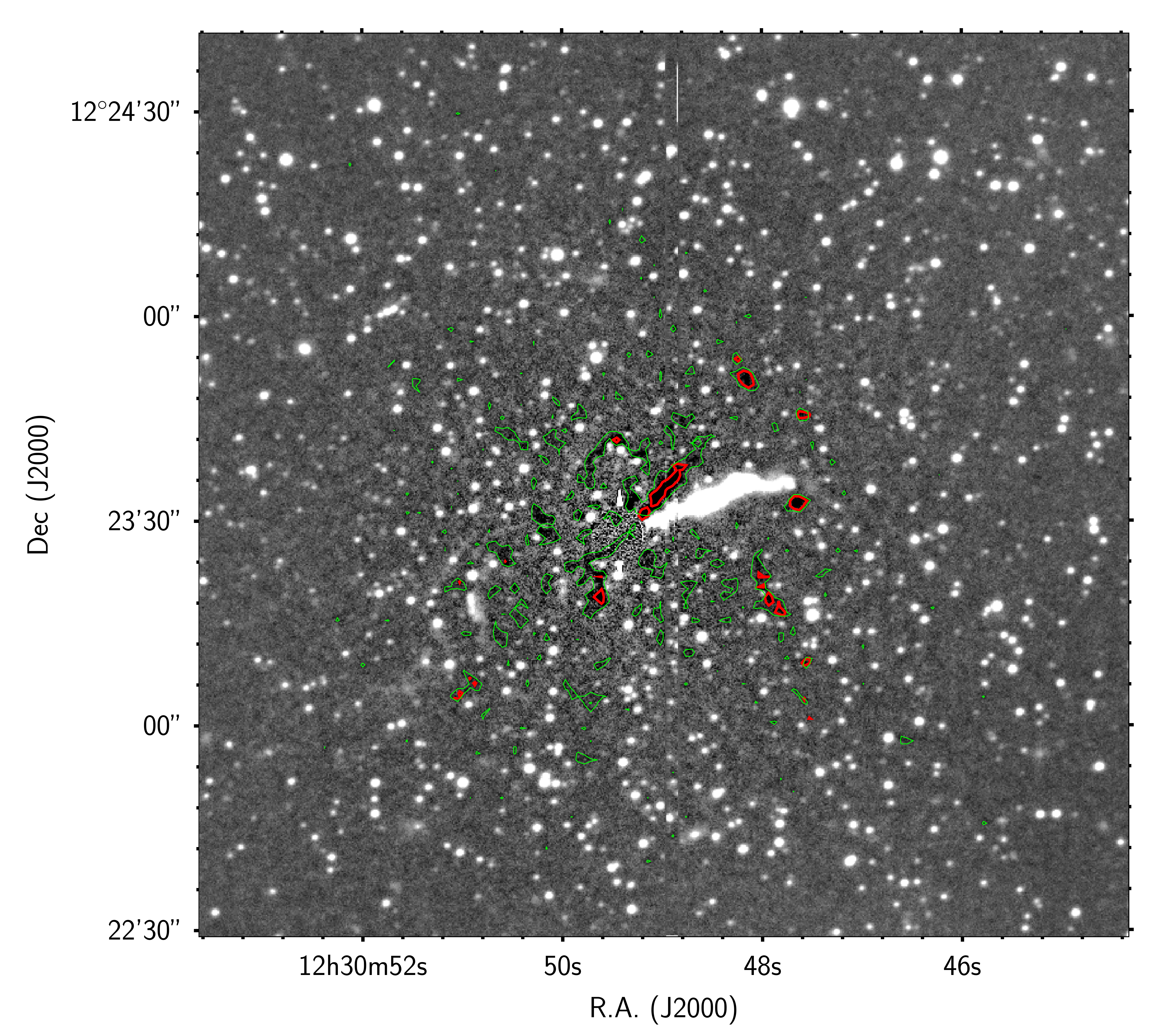}
   \caption{The stellar continuum subtracted $g$-band image of M87. The dark regions outlined by green contours are dusty structures
   detected in absorption. The red contour corresponds to $A_g$ = 0.010$\pm$ 0.002.
   The bright elongated region is the jet and the point sources globular clusters and foreground stars.
 }
   \label{NGVS}%
   \end{figure}

The availability of deep and extended surveys targeting M87, allows us to investigate whether dust is also present at larger distances from the galaxy's center. 
To do so we used the $g$-band imaging gathered by the NGVS, that in the galaxy central regions reaches optimal seeing conditions, 0.62\arcsec, and a depth 
of $g = 27.3$ mag or $\mu_{g} = 29.0 $ mag arcsec$^{-2}$. The continuum emission from the
galaxy was retrieved using the IRAF tasks ISOFIT and C-MODEL (Ciambur et al. 2015) that allow for an accurate recovery of the 
galaxy surface brightness profile, hence of a two-dimensional model of the galaxy light. The resulting stellar continuum subtracted image of the inner 
$2 \arcmin\times2\, \arcmin$ is shown in Fig. \ref{NGVS}. Despite the poor image quality in the central 0.05\arcmin$\times$0.05\arcmin, affected by the 
presence of the AGN (see Bellini et al. 2015 for the detection of dust in this region), Fig. \ref{NGVS} clearly shows the presence of dust in 
absorption. This is detected along the filament located to the north of the jet and in several other patchy features close to the centre as already identified 
by previous analyses (Sparks et al. 1993, Ferrarese et al. 2006, Bellini et al. 2015). However, the high quality NGVS imaging also reveals a 
complicated network of dusty regions unknown so far. These are distributed around the galaxy in the western direction up to $\sim$2.4 kpc (0.5$^{\prime}$). 

At its highest densities the most prominent dust feature (red contours in Fig. \ref{NGVS}) absorbs the continuum light with an 
extinction factor $A_{g} = 0.010\pm 0.002 - 0.017\pm 0.002$ mag. Such a value was estimated by comparing the NGVS $g$-band image with a dust-free model 
of the light distribution, the latter given by the ISOFIT two-dimensional map of the galaxy light. We note that, as consequence of 
the relatively small covering area of the dust, the ISOFIT model is a good approximation of the unextinguished light 
because the result of an azimuthal average. 
We also remark that this extinction factor perfectly agrees with the value obtained by 
Sparks et al. (1993). As in this work, assuming a standard gas-to-dust ratio $d$ $\simeq$ 100 and the relation:

\begin{equation}
{N(H) = 1.8 \times 10^{23} \frac{A_{V}}{d}  [\rm{cm^{-2}}]}
\end{equation}

\noindent 
we can calculate the expected column density of the cold gas associated with the dust lane, $N(H)$ $\simeq$ 1.8 $\times$ 10$^{19}$ cm$^{-2}$.
This number is consistent with the upper limit in the column density derived from observations of HI in absorption (van Gorkom et al. 1989, 
Dwarakanath et al. 1994) or from other absorption line measurements and from the shape of the X-rays spectrum in the nucleus (Sabra et al. 2003).
Given the size of the filament, we estimate as in Sparks et al. (1993) that the associated total mass of cold gas is 
$M_{gas,dust}$ $\simeq$ 2 $\times$ 10$^7$ M$_{\odot}$. 

\section{Analysis}

\subsection{Narrow-band imaging}


The continuum-subtracted narrow-band image of M87 shows prominent ionised gas filaments extending from the nucleus along an axis with a position angle of $\sim$ 55 deg
measured clockwise from the north. Most of the extended emission is on the south-eastern side of the galaxy, where the emitting gas is detected up to $\sim$ 8 kpc from the
nucleus, while only $\sim$ 3 kpc in the north-western direction. The typical surface density of the gas here is $\Sigma(H\alpha+[NII])$ $\simeq$ 2.5$\times$10$^{-17}$
erg s$^{-1}$ cm$^{-2}$ arcsec$^{-2}$ in the diffuse features while exceeding $\Sigma(H\alpha+[NII])$ = 10$^{-16}$ erg s$^{-1}$ cm$^{-2}$ arcsec$^{-2}$ in 
the brightest clumpy regions. 
The total flux of the main central filament, measured within a rectangular box 133$\times$35 arcsec inclined at 55 deg from north (clockwise), is 
$f(H\alpha+[NII])$ = 1.036 $\times$ 10$^{-12}$ erg s$^{-1}$ cm$^{-2}$. Assuming that the typical [NII] contamination is log[NII]$_{6583}$/H$\alpha$ $\simeq$ 0.25 
as measured within the MUSE field (see Fig. \ref{shock}), its total H$\alpha$ luminosity is $L(H\alpha)$ = 9.56 $\times$ 10$^{39}$ erg s$^{-1}$. 
As in Fossati et al. (2016) and Boselli et al. (2016) we can estimate the mean density and the total mass of the gas emitting in H$\alpha$ along the filament using the relation:

\begin{equation}
{L(H\alpha) = n_e n_p \alpha^{eff}_{H\alpha} V f h \nu_{H\alpha}}
\end{equation}

\noindent
(Osterbrock \& Ferland 2006), where $n_e$ and $n_p$ are the number density of electrons and protons, $\alpha^{eff}_{H\alpha}$
is the H$\alpha$ effective recombination coefficient ($\alpha^{eff}_{H\alpha}$ = 1.17 $\times$10$^{-13}$), $V$ is the volume of the emitting region, $f$ the volume filling factor, 
$h$ the Planck's constant, and $\nu_{H\alpha}$ the frequency of the H$\alpha$ transition. Assuming a cylindrical 
distribution of the gas, with an orientation of 45 deg along the line of sight, and a filling factor $f$ = 0.1,  
we estimate that the typical density of the gas is $n_e$ $\simeq$ 0.3
cm$^{-3}$ and its total mass $M(ionised)_{filament}$ $\simeq$ 6.9 $\times$ 10$^7$ M$_{\odot}$. We stress, however, that this estimate, which is $\simeq$ a factor of 10 larger than the
mass derived by Sparks et al. (1993), is a very rough estimate given the very high uncertainty on the filling factor of
the gas and on the volume of the filament. 

\begin{figure}
   \centering
   \includegraphics[width=0.5\textwidth]{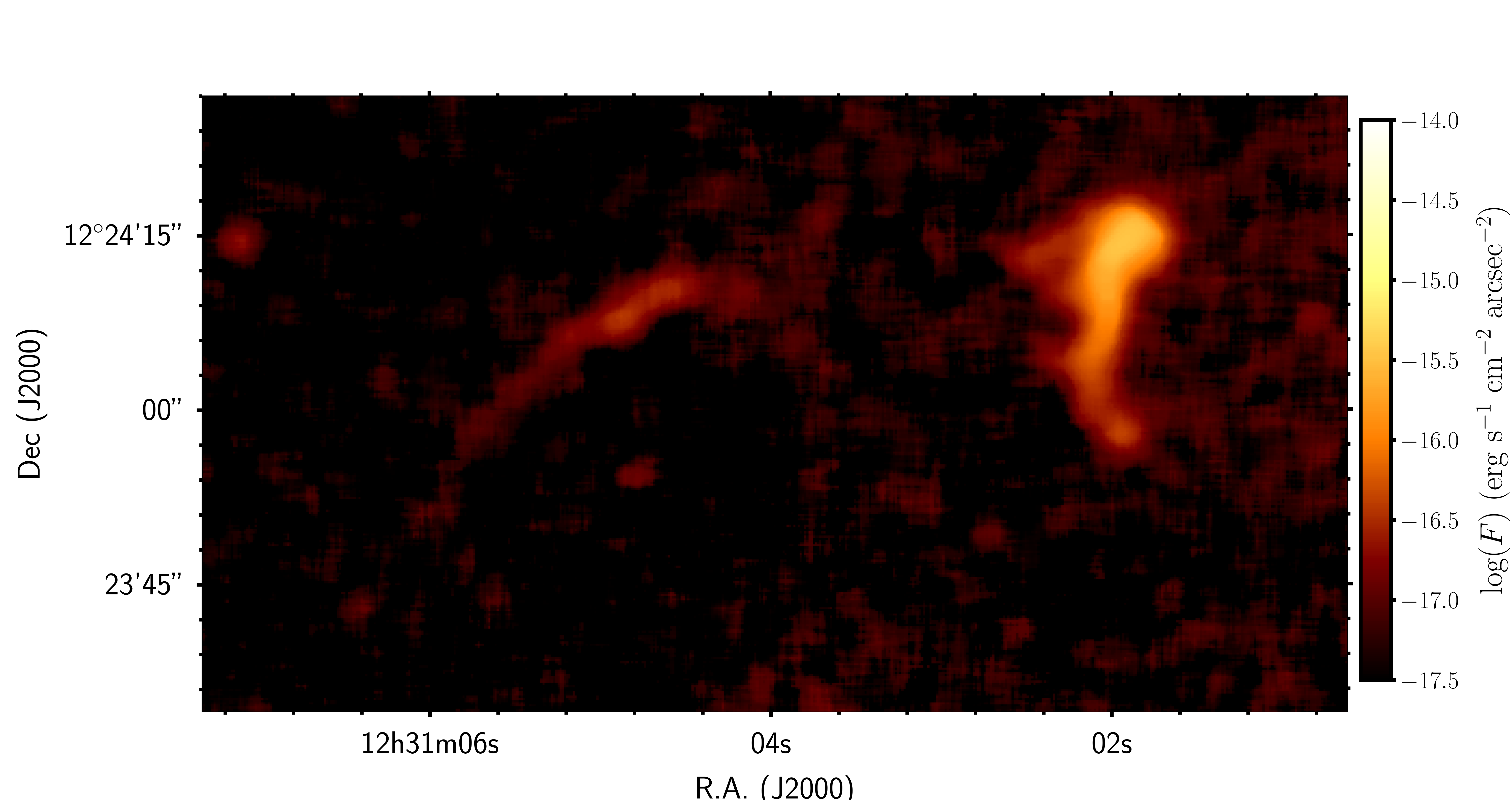}
   \caption{Zoom on the plume of ionised gas discovered by Gavazzi et al. (2000) (west), and on a new filament of ionised gas $\simeq$ 3 kpc long (east).
   This zoomed region is indicated with a dashed green box in Fig. \ref{M87Hasmo}. }
   \label{M87plume}%
   \end{figure}

The plume of ionised gas at $\sim$ 15 kpc to the east discovered by Gavazzi et al. (2000) is also evident given its high surface brightness
($\Sigma(H\alpha+[NII])$ $\simeq$ 10$^{-16}$ erg s$^{-1}$ cm$^{-2}$ arcsec$^{-2}$, see Fig. \ref{M87plume}). Figure \ref{M87plume}
also shows a previously-unknown banana-shaped, low surface brightness feature 
($\Sigma(H\alpha+[NII])$ = 5$\times$10$^{-18}$ erg s$^{-1}$ cm$^{-2}$ arcsec$^{-2}$) $\simeq$ 3 kpc long
at $\sim$ 3 kpc east from this plume (18 kpc from the galaxy nucleus).


\subsection{Spectroscopy}

\subsubsection{Gas kinematics}

\begin{figure*}
   \centering
   \includegraphics[width=1\textwidth]{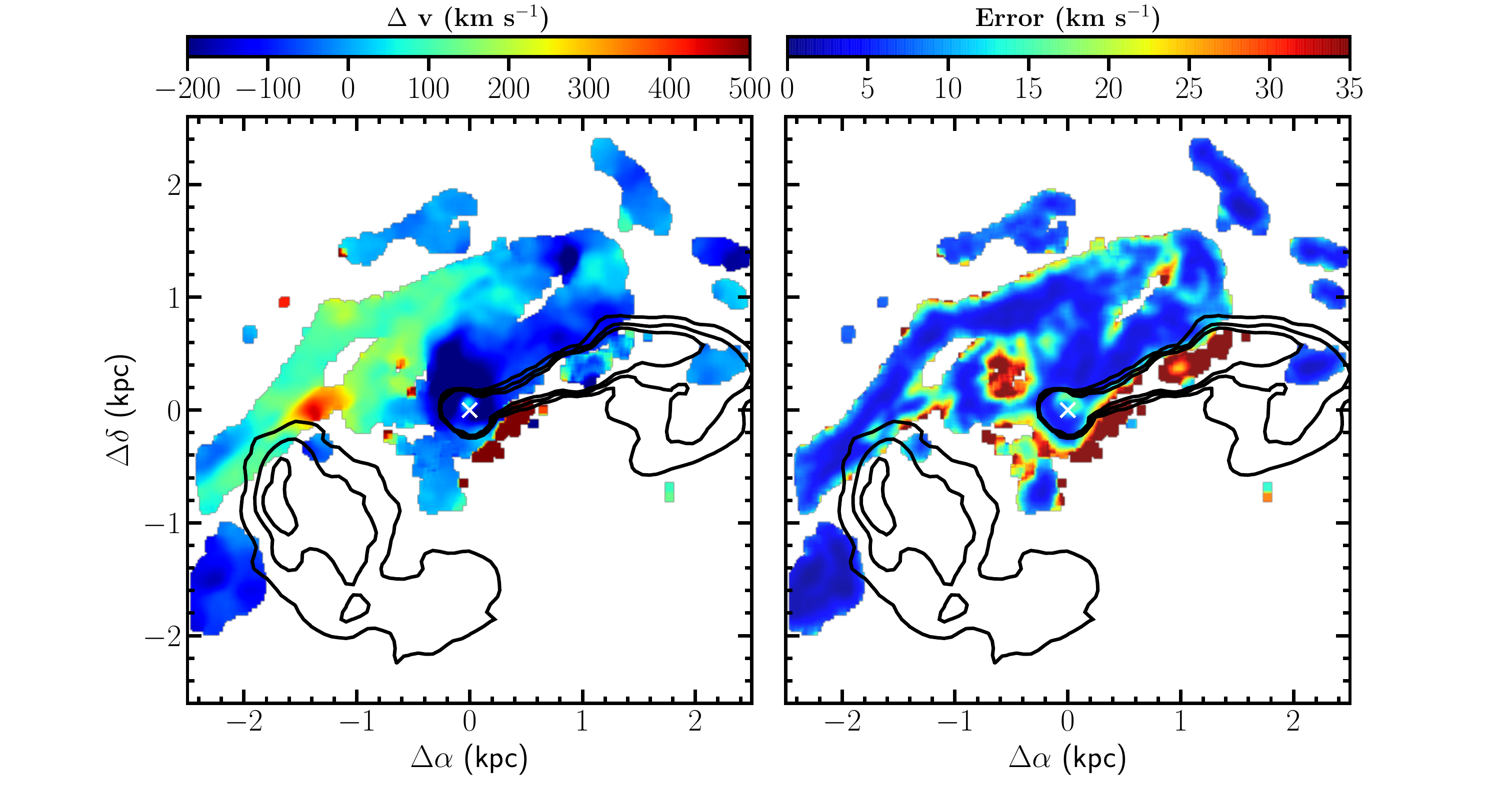}
   \caption{Velocity map (left) and associated error (right) of the ionised gas derived from MUSE for emission lines with a $S/N$ $>$ 10. The velocity of the gas is given relative to the systemic velocity of
   the galaxy of 1292 km s$^{-1}$. 
 }
   \label{vel}%
   \end{figure*}
\begin{figure*}
   \centering
   \includegraphics[width=1\textwidth]{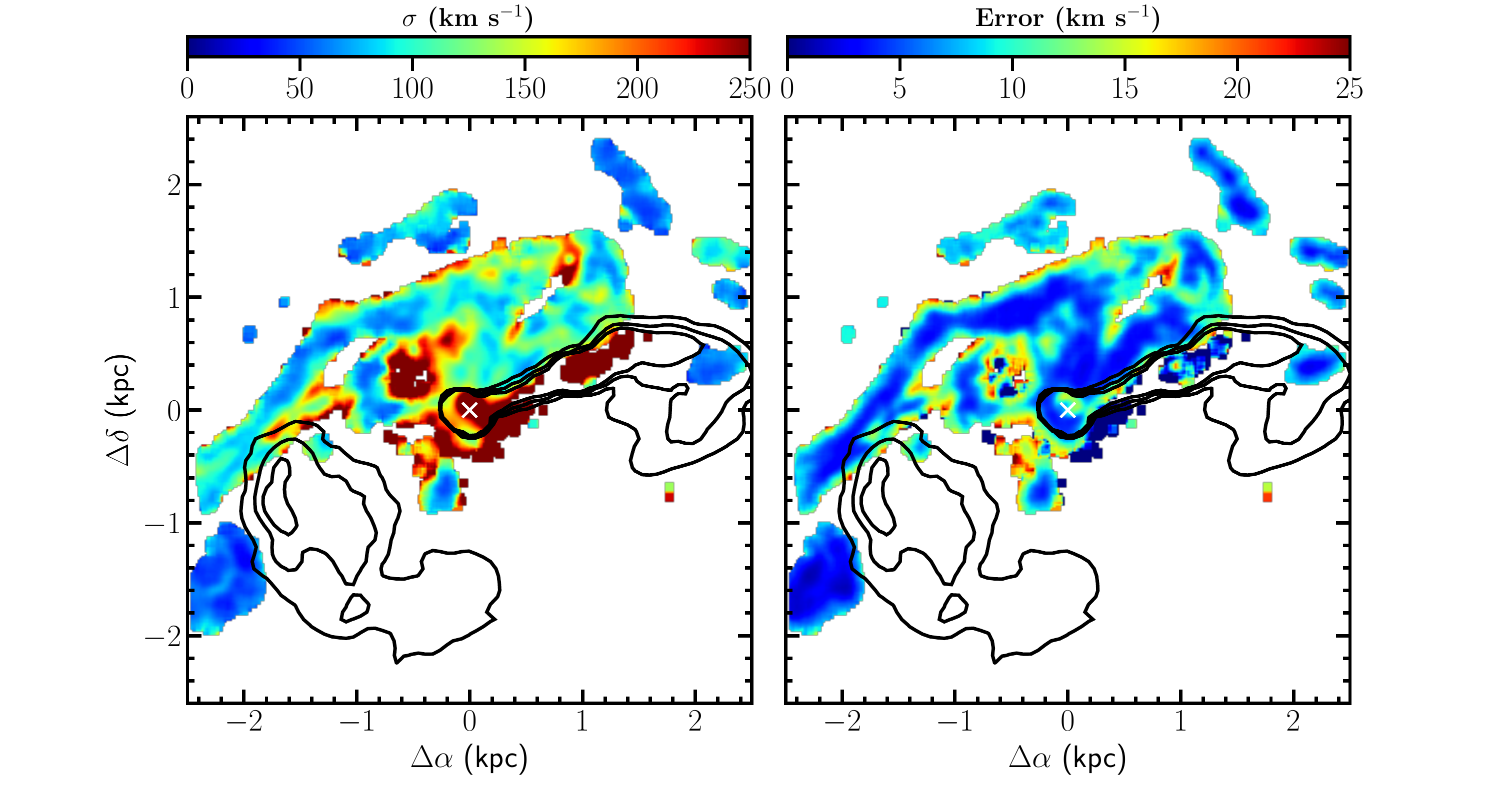}
   \caption{Velocity dispersion map (left) and associated error (right) of the ionised gas derived from MUSE for emission lines with a $S/N$ $>$ 10. 
 }
   \label{sigma}%
   \end{figure*}

Figure \ref{vel} shows the velocity of the gas along the filament within the central 1\arcmin$\times$1\arcmin region. Thanks to the IFU mode of MUSE, the 
improvement in quality with respect to previous long slit data is astonishing (Heckman et al. 1989, Sparks et al. 1993, Werner et al. 2013). Figure \ref{sigma} shows the map 
of the velocity dispersion of the gas. The gas does not follow any ordered motion. As already noticed by Sparks et al. (1993), the difference in velocity 
within the inner gas filament is very high, reaching 700-800 km s$^{-1}$ on scales $\lesssim$ 1 kpc. On the contrary, the velocity dispersion is $\simeq$ 100 km s$^{-1}$
and is fairly uniform over the gas filament, with the only exception on the nucleus where it reaches $\simeq$ 450 km s$^{-1}$ because of the presence of an AGN. 
These velocity dispersions are generally larger than those measured within the tails of ionised gas observed in other perturbed cluster galaxies 
where these values are reached only within limited regions well outside the optical disc (Fumagalli et al. 2014, Consolandi et al. 2017, Poggianti et al. 2017,
Bellhouse et al. 2017). For comparison, within the same MUSE field the velocity dispersion of the stellar component goes from $\simeq$ 370 km s$^{-1}$ in the central
2 arcsec (160 pc) to $\simeq$ 260 km s$^{-1}$ at $\sim$ 1.6 kpc, whereas the velocity rotation is always $\lesssim$ 10 km s$^{-1}$ (Emsellem et al. 2014).
The steep gradient in the velocity field observed in the regions at (-1.4; 0.0) kpc from the nucleus (Fig. \ref{vel}, left panel), 
which is also characterised by a high velocity dispersion (Fig. \ref{sigma}, left panel) (and a high uncertainty on both the rotational velocity and the
velocity dispersion, Figs. \ref{vel} and \ref{sigma}, right panels) suggests the presence of two physically and kinematically distinct components in the ionised gas.
Cold gas has been detected in the bright ionised gas region at the south-east edge of the counter-jet radio lobe at (-2.2; -1.5) kpc from the nucleus 
through the emission of the $^{12}$CO(2-1) (Simonescu et al. 2018) and [CII] (Werner et al. 2013) lines. Their recessional velocity is consistent within the errors.

Sparks et al. (1993), using geometrical arguments based on the presence of a dust lane absorbing the stellar emission, 
concluded that the associated gas filament in the western direction is in the foreground, and thus it is flowing out from the nucleus.
Using similar arguments based on the 
relative position of the ionised gas filament and of the radio lobe, combined with the lack of any dust component, they also concluded that the south-eastern gas filament
is in the background of the galaxy.

\subsubsection{Emission line properties}

\begin{figure}
   \centering
   \includegraphics[width=0.5\textwidth]{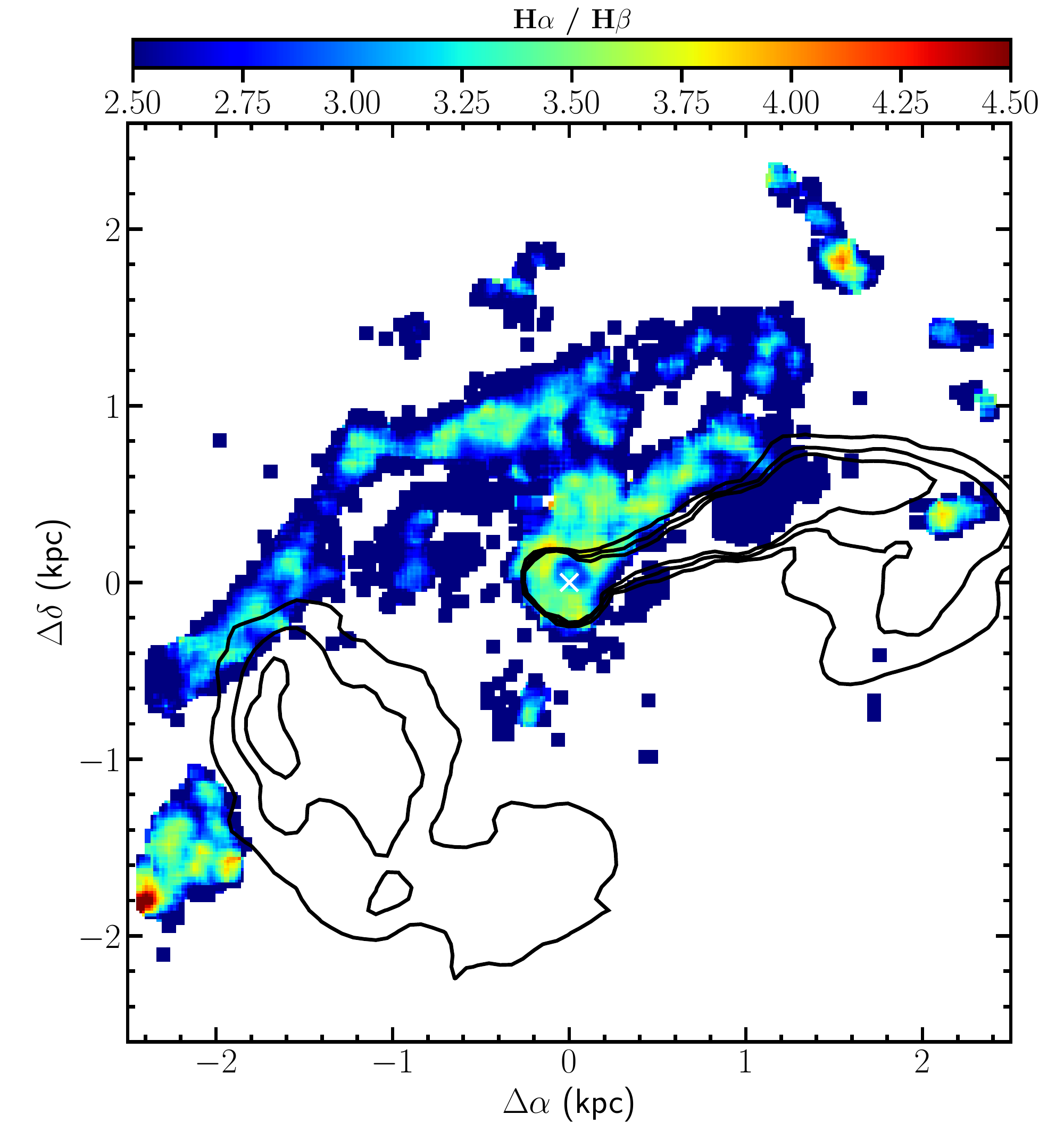}
   \caption{Distribution of the Balmer decrement derived for emission lines with a $S/N$ $>$5.
 }
   \label{BD}%
   \end{figure}

Several gas emission lines in the MUSE spectral range (4800-9000 \AA) can be used to derive 
the physical properties of the gas through diagnostic diagrams. The H$\alpha$/H$\beta$ Balmer decrement is generally used to 
derive the dust attenuation within the ionised gas. 
The Balmer decrement within the MUSE field is shown in Fig. \ref{BD}. The H$\alpha$/H$\beta$ ratio is fairly constant along the filament, 
with values close to H$\alpha$/H$\beta$ $\simeq$ 2.8-3.5. These values are close to those expected whenever the gas is photoionised by the emission
of young and massive stars and the dust content is low ($E(B-V)$ $\lesssim$ 0.08). Similar values, however, are also expected whenever the gas 
is collisionally-heated by cosmic rays accelerated by magnetohydrodynamic waves in the filament (H$\alpha$/H$\beta$ $\simeq$ 3.7-4.4; Ferland et al. 2009).
The comparison of Fig. \ref{BD} and Fig. \ref{NGVS} gives a consistent picture where dust attenuation is at the origin of the observed Balmer decrement.

\begin{figure}
   \centering
   \includegraphics[width=0.5\textwidth]{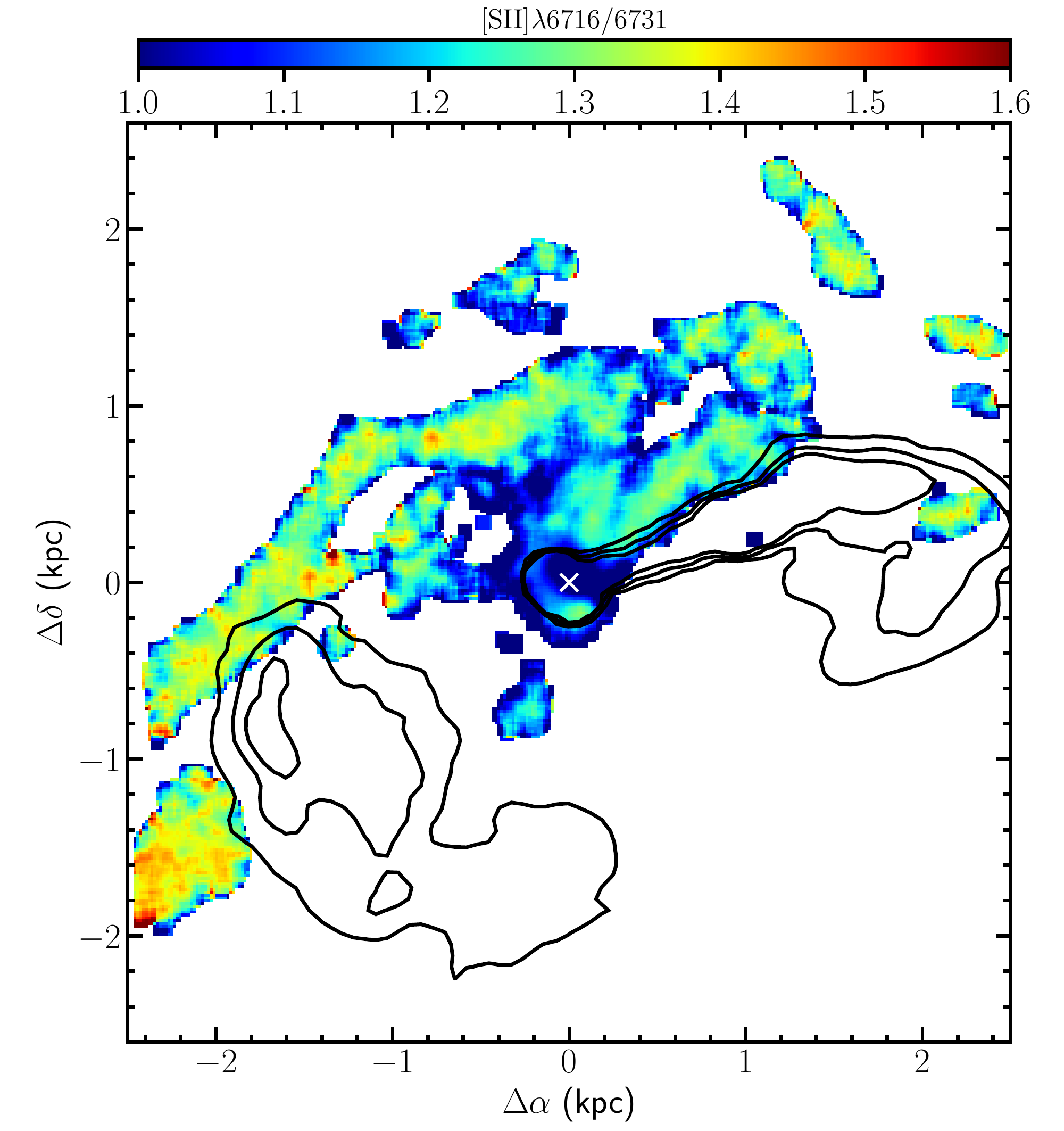}
   \caption{Distribution of the [SII]$\lambda$6716\AA/[SII]$\lambda$6731\AA ~ ratio for emission lines with a $S/N$ $>$5. 
 }
   \label{SII}%
   \end{figure}

The [SII]$\lambda$6716\AA/[SII]$\lambda$6731\AA, on the other hand, is tightly connected to the density of the gas (Osterbrock \& Ferland 1989).
Figure \ref{SII} shows that, within the inner 1\arcmin$\times$1\arcmin region of M87 mapped by MUSE the sulphur doublet ratio
smoothly increases from [SII]$\lambda$6716\AA/[SII]$\lambda$6731\AA $\simeq$ 0.9 ($n_e$ $\simeq$ 6 $\times$ 10$^2$ cm$^{-3}$ assuming a temperature of 10$^4$ K
and using the recent calibration of Proxauf et al. 2014) 
close to the nucleus to [SII]$\lambda$6716\AA/[SII]$\lambda$6731\AA $\simeq$ 1.35 ($n_e$ $\simeq$ 80 cm$^{-3}$) 
at the edges of the field\footnote{These densities scales as (10$^4$/$T$)$^{-1/2}$ if the temperature is $T$ $\neq$ 10$^4$ K, Osterbrock \& Ferland (2006).}. These low densities are similar to those
encountered in other tails of ionised gas in perturbed cluster galaxies (e.g. Poggianti et al. 2017). We notice that a large 
[SII]$\lambda$6716\AA/[SII]$\lambda$6731\AA $\geq$ 1.4 ratio has been also measured by Gavazzi et al. (2000) in the plume of ionised gas at $\sim$ 15 kpc to the east.

Unfortunately the high noise induced by the bright stellar continuum in these inner regions prevents the detection of 
the weak [NII] line at $\lambda$ = 5755 \AA, we are thus not able to estimate the variation of the gas temperature along the filament using the 
[NII]$\lambda$6548 + [NII]$\lambda$6583\AA/[NII]$\lambda$5755\AA ~line ratio (Osterbrock \& Ferland 2006). 
The detection of the [CIV] line at $\lambda$ 1549 \AA ~ by Sparks et al. (2012) in the ionised gas, however, suggests that the gas might reach temperatures
of $T$ = 10$^5$ K. At this temperature, the iron lines [FeXIV] at $\lambda$ 5303 \AA ~ and [FeX] at $\lambda$ 6374 \AA ~should also be present (Heckman et al. 1989). As explained above, however, the
detection of these weak emission lines is hampered by the bright stellar continuum.

\begin{figure*}
   \centering
   \includegraphics[width=1\textwidth]{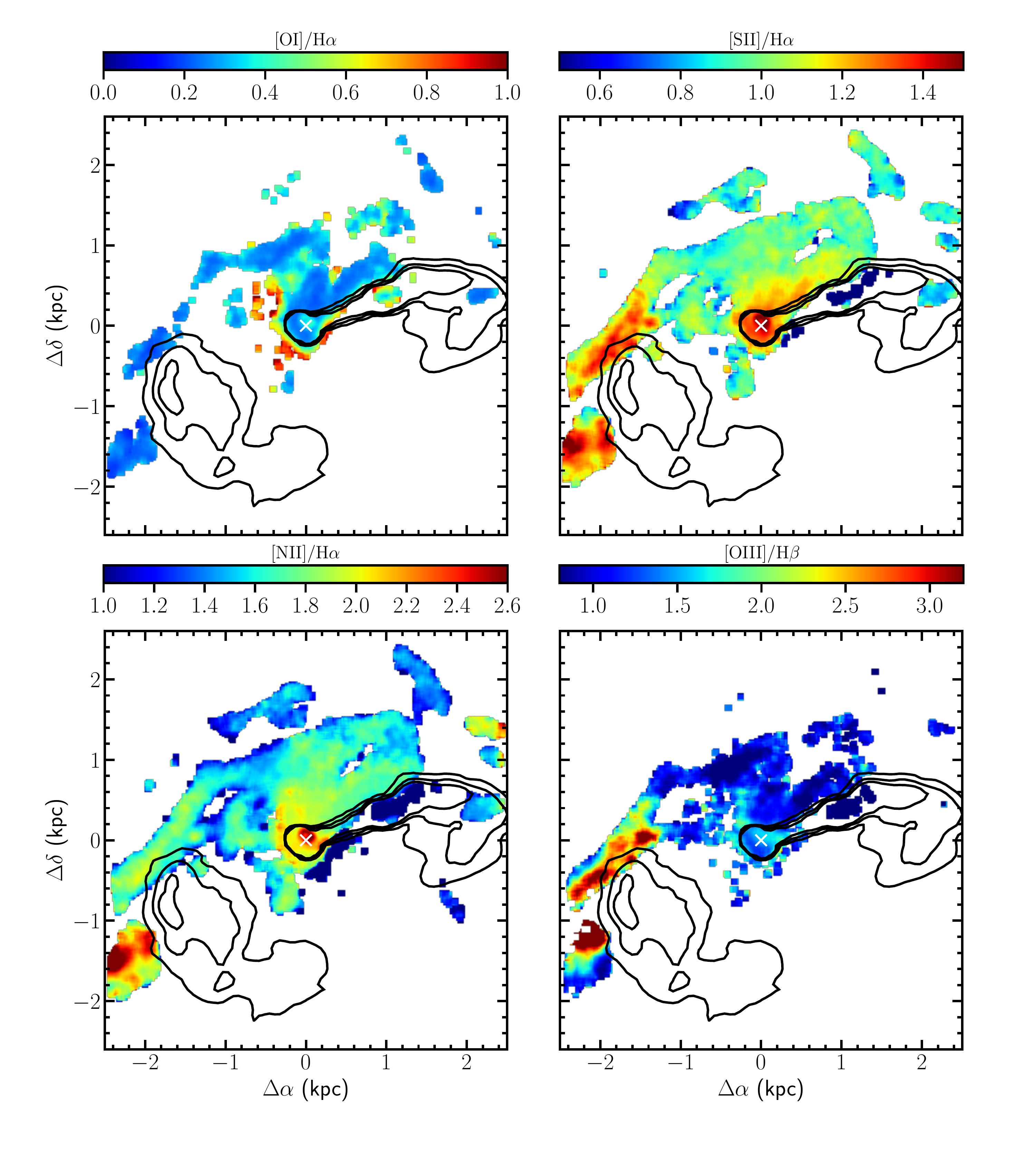}
   \caption{Distribution of the line ratios [OI]$\lambda$6300\AA/H$\alpha$ (upper left), [SII]$\lambda$6716,6731\AA/H$\alpha$ (upper right), 
   [NII]$\lambda$6548,6583\AA/H$\alpha$ (lower left), and [OIII]$\lambda$5007\AA/H$\alpha$ line ratios for emission lines with a $S/N$ $>$5.   
 }
   \label{shock}%
   \end{figure*}

Figure \ref{shock} shows the map of different line ratios within the MUSE field, including the shock-sensitive [OI]$\lambda$6300\AA/H$\alpha$, 
[SII]$\lambda$6716,6731\AA/H$\alpha$ and [NII]$\lambda$6548,6583\AA/H$\alpha$
line ratios (Rich et al. 2011). Figure \ref{shock} shows high values for all the shock-sensitive tracers, both close to the AGN or along the filament, suggesting that 
within these regions the gas is mainly ionised by shocks. 
For comparison, lower values ([OI]$\lambda$6300\AA/H$\alpha$ =0.18; [SII]$\lambda$6716,6731\AA/H$\alpha$=0.53) have been 
measured in the plume of ionised gas discovered by Gavazzi et al. (2000) at 15 kpc east from the galaxy nucleus.

\begin{figure*}
   \centering
   \includegraphics[width=0.8\textwidth]{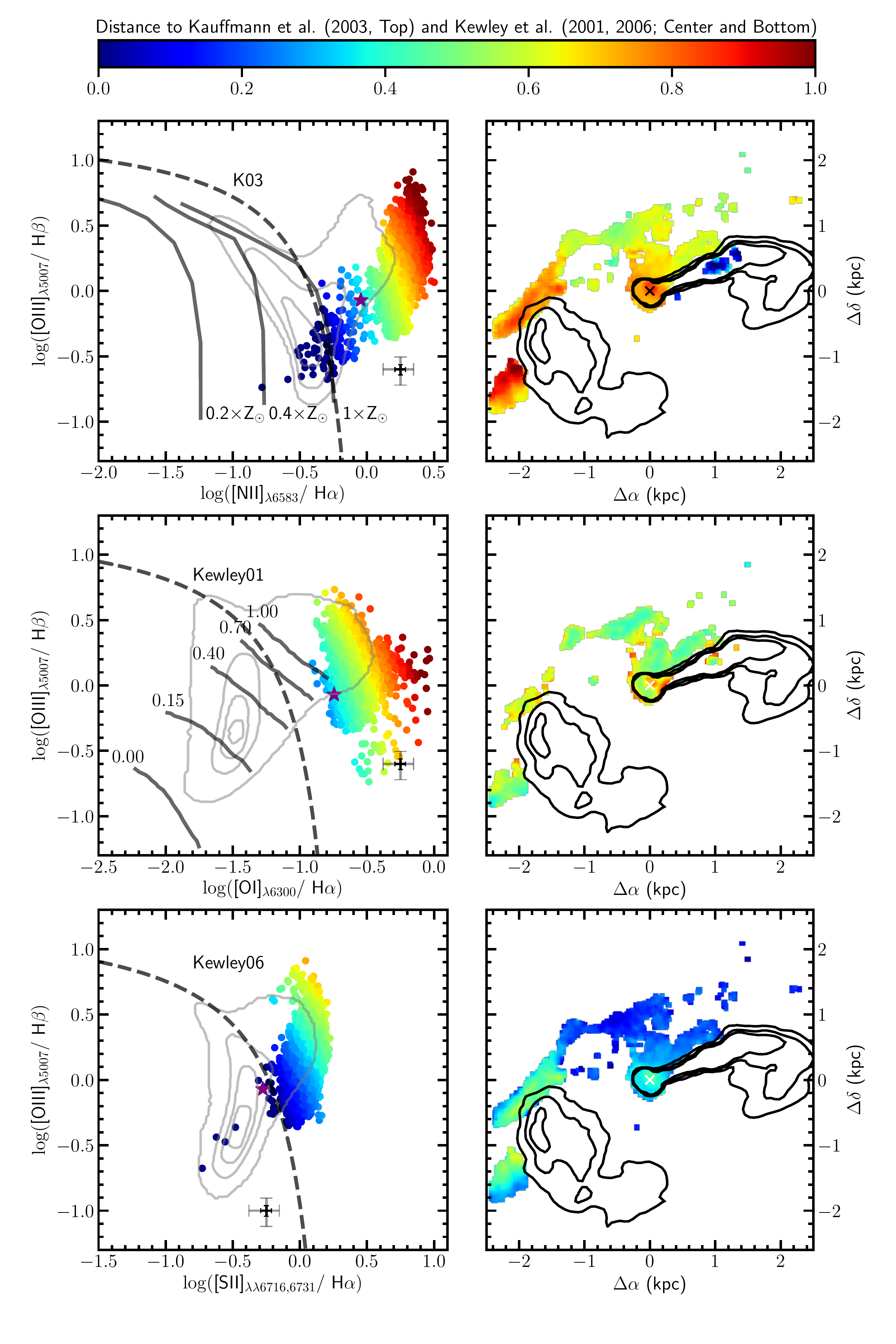}
   \caption{Left: [OIII]/H$\beta$ vs. [NII]/H$\alpha$ (top), [OIII]/H$\beta$ vs. [OI]/H$\alpha$ (centre), and [OIII]/H$\beta$ vs. [SII]/H$\alpha$ (bottom) BPT diagrams for emission lines with a $S/N$ $>$5. 
   The dashed curves separate AGN from HII regions (Kauffmann et al. 2003; Kewley et al. 2001, Kewley et al. 2006). 
   Data are colour coded according to their minimum distance 
   from these curves. The black and gray crosses indicate the typical error on the data for lines with S/N $\simeq$ 15 and S/N $\simeq$ 5, respectively.
   The gray contours show the distribution of a random sample of nuclear spectra of SDSS galaxies in the redshift range 0.01 - 0.1 and 
   stellar masses 10$^9$ $\leq$ $M_{star}$ $\leq$ 10$^{11}$ M$_{\odot}$.
   The magenta star shows the position in the diagram of the plume of ionised gas observed at 15 kpc to the east by Gavazzi et al. (2000). 
   The thick solid lines in the upper left panel show three different photo-ionisation models at different metallicities (0.2, 0.4, 1 Z$_{\odot}$; Kewley et al. 2001),
   those in the middle left panel the shock models of Rich et al. (2011) for increasing shock fractions (from left to right) in a twice solar gas.
   Right: map of the spaxel distribution colour-coded according to their position in the BPT diagram. 
   }
   \label{BPT}%
   \end{figure*}

Figure \ref{BPT} shows three typical line diagnostic diagrams generally referred to as BPT diagrams (Baldwin et al. 1981). Thanks to the IFU mode of MUSE and
to its excellent sensitivity, the difference with respect to published BPT diagrams all based on long-slit spectroscopy (Heckman et al. 1989, Crawford et al. 1999), is stunning. 
The three BPT diagrams consistently indicate that the gas filament
is not photoionised by massive young stars, in agreement with the lack of young stellar associations in the UV images of the galaxy. 
All the points, indeed, are located to the upper right of the typical curves delimiting the position of HII regions from 
AGN. The handful of dots in the photoionisation regime in the upper and lower panels of Fig. \ref{BPT} are located along the jet, thus their line ratio estimates 
is highly uncertain. 

\begin{figure*}
   \centering
   \includegraphics[width=1\textwidth]{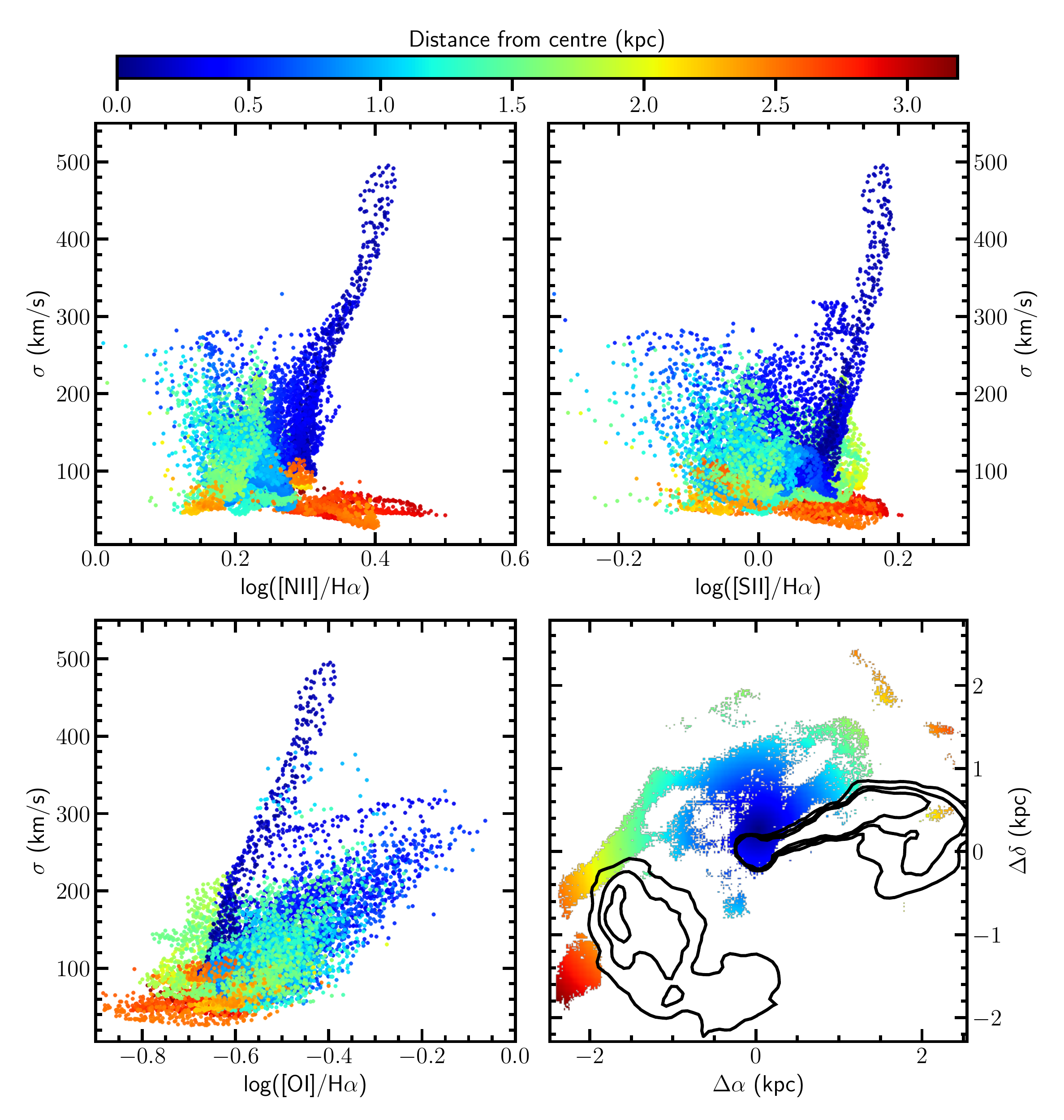}
   \caption{Relationship between the velocity dispersion in the ionsied gas and the shock-sensitive line ratios [NII]$\lambda$6584/H$\alpha$ (upper left),
   [SII]$\lambda$6716,6731\AA/H$\alpha$ (upper right), and
   [OI]$\lambda$6300\AA/H$\alpha$ (lower left) measured for emission lines with a $S/N$ $>$ 3 (and $S/N$ $>$ 10 for the determination of the velocity dispersion
   as described in sect. 2.2). Each point is colour-coded according to its distance from the galaxy centre, as indicated in the right lower panel.
   }
   \label{sigma2}%
   \end{figure*}

To identify the possible ionising source of the gas within the filament we plot in Fig. \ref{sigma2} the relationship between the velocity dispersion of the gas $\sigma$
and the three different shock-sensitive line ratios [NII]$\lambda$6584/H$\alpha$, [SII]$\lambda$6716,6731\AA/H$\alpha$, and [OI]$\lambda$6300\AA/H$\alpha$. 
Points are colour-coded according to their distance from the galaxy centre. 
A strong relationship between 
the two variables has been interpreted in the literature as a further indication that the gas is excited by shocks (Rich et al. 2011, 2015, Ho et al. 2014).
A significant correlation is seen only in the [OI]$\lambda$6300\AA/H$\alpha$ line ratio, 
consistently with what found in previous works (Ho et al.
2014). Here the velocity dispersion of the gas increases with the shock-sensitive [OI]$\lambda$6300\AA/H$\alpha$ line ratio roughly in two different ways:
there is a very tight and steep relation spanning the whole range in velocity dispersion (0 $\lesssim$ $vel$ $\lesssim$ 500 km s$^{-1}$) where the line ratio increases from 
log([OI]$\lambda$6300\AA/H$\alpha$) $\simeq$ -0.7 to log([OI]$\lambda$6300\AA/H$\alpha$) $\simeq$ -0.4. The dark-blue colour of the points suggests that this tight 
relation is due to the inner regions close to the AGN. The bulk of the points forms a flatter and much more dispersed relation spanning the whole range in line ratio
(-0.8 $\lesssim$ log([OI]$\lambda$6300\AA/H$\alpha$) $\lesssim$ 0.3) but reaching only velocity dispersions up to $\sigma$ $\simeq$ 300 km s$^{-1}$. This trend is due to 
the main body of the filament (light-blue, cyan, and green points). 
The regions located at the edge of the counter-jet (red points) do not show any relation.
The tight and steep relation in the inner region (dark-blue points) is also clear in the $\sigma$ vs. [NII]$\lambda$6584/H$\alpha$ and [SII]$\lambda$6716,6731\AA/H$\alpha$ line ratios diagrams. 
In these other two diagrams, however, the bulk of the points does not follow the $\sigma$ vs. line ratio relation observed in the [OI] line. In particular, we remark that
those located at the south-east edge of the filament observed by MUSE, at $\simeq$ 2 kpc from the nucleus of the galaxy (red points), have the lowest velocity dispersions
but the highest [NII]$\lambda$6584/H$\alpha$ and [SII]$\lambda$6716,6731\AA/H$\alpha$ line ratios within the MUSE field.


\subsection{Comparison with other bands}

\begin{figure*}
   \centering
   \includegraphics[width=1\textwidth]{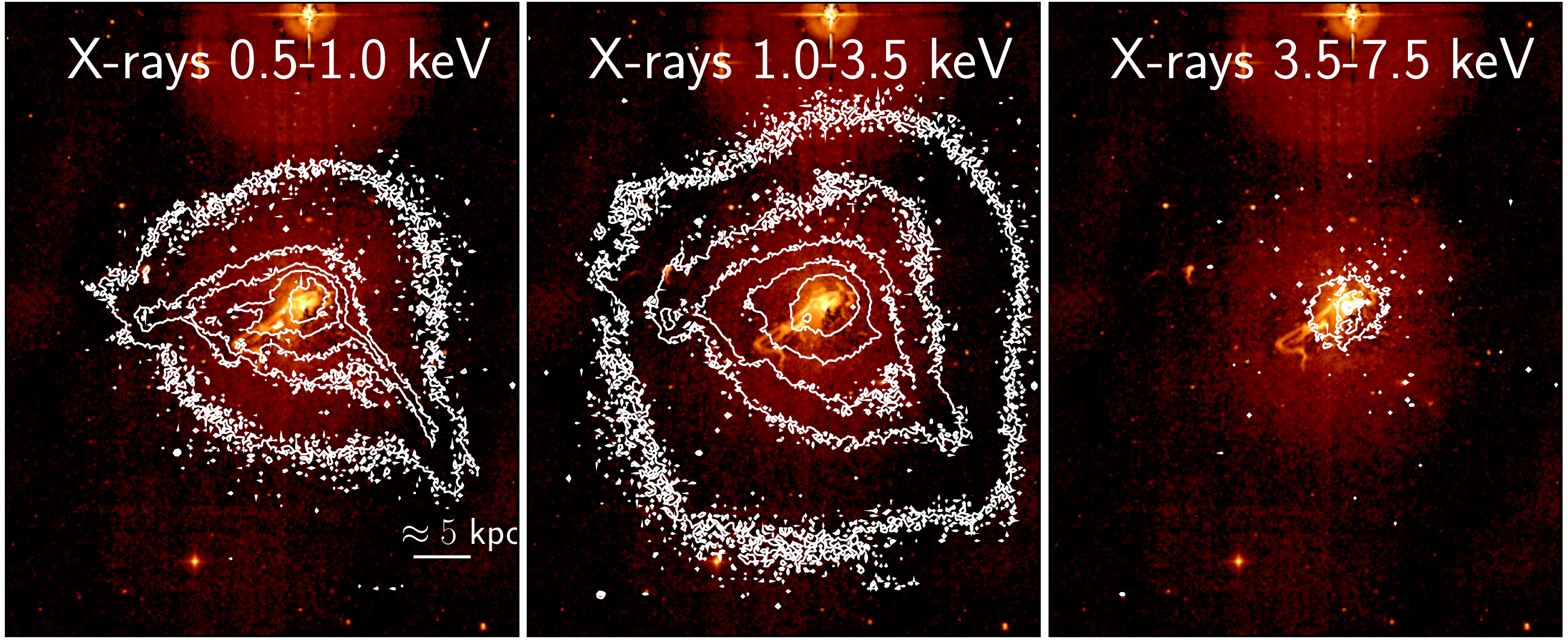}
   \caption{Comparison between the H$\alpha$+[NII] emission (in colour) and the X-rays emission (contours) in the three bands 0.5-1.0 keV (left), 
   1.0-3.5 keV (centre), and 3.5-7.0 keV in white contours (from Forman et al. 2007).
 }
   \label{M87HaX}%
   \end{figure*}

The spectacular images at hand allow us to make a detailed comparison between the distribution of the ionised gas 
emission and the other components of the interstellar medium of the galaxy. Figure \ref{M87HaX} shows with a sensitivity and resolution never
reached so far (Sparks et al. 1993, 2004, Werner et al. 2013) how the hot gas traced by the X-rays emission from \textit{Chandra} compares 
with the ionised gas emission. As noticed in previous works (Sparks et al. 1993, 2004, Werner et al. 2013), at large scale 
the main ionised gas filaments are not associated with the hot gas distribution which has a structured shape only in the 0.5-1.0 and 1.0-3.5 keV bands,
while at smaller scales the coincidence between well defined features is present (e.g. Sparks et al. 2004). 
We remark that the plume of ionised gas first detected by Gavazzi et al. (2000)
at $\simeq$ 15 kpc to the east, and the extended low surface brightness filament $\simeq$ 3 kpc to the east of this plume, at $\simeq$ 18 kpc from the nucleus, 
first detected in this work, correspond to sharp edges in the X emission in the 0.5-1.0 and 1.0-3.5 keV bands. 

\begin{figure}
   \centering
   \includegraphics[width=0.5\textwidth]{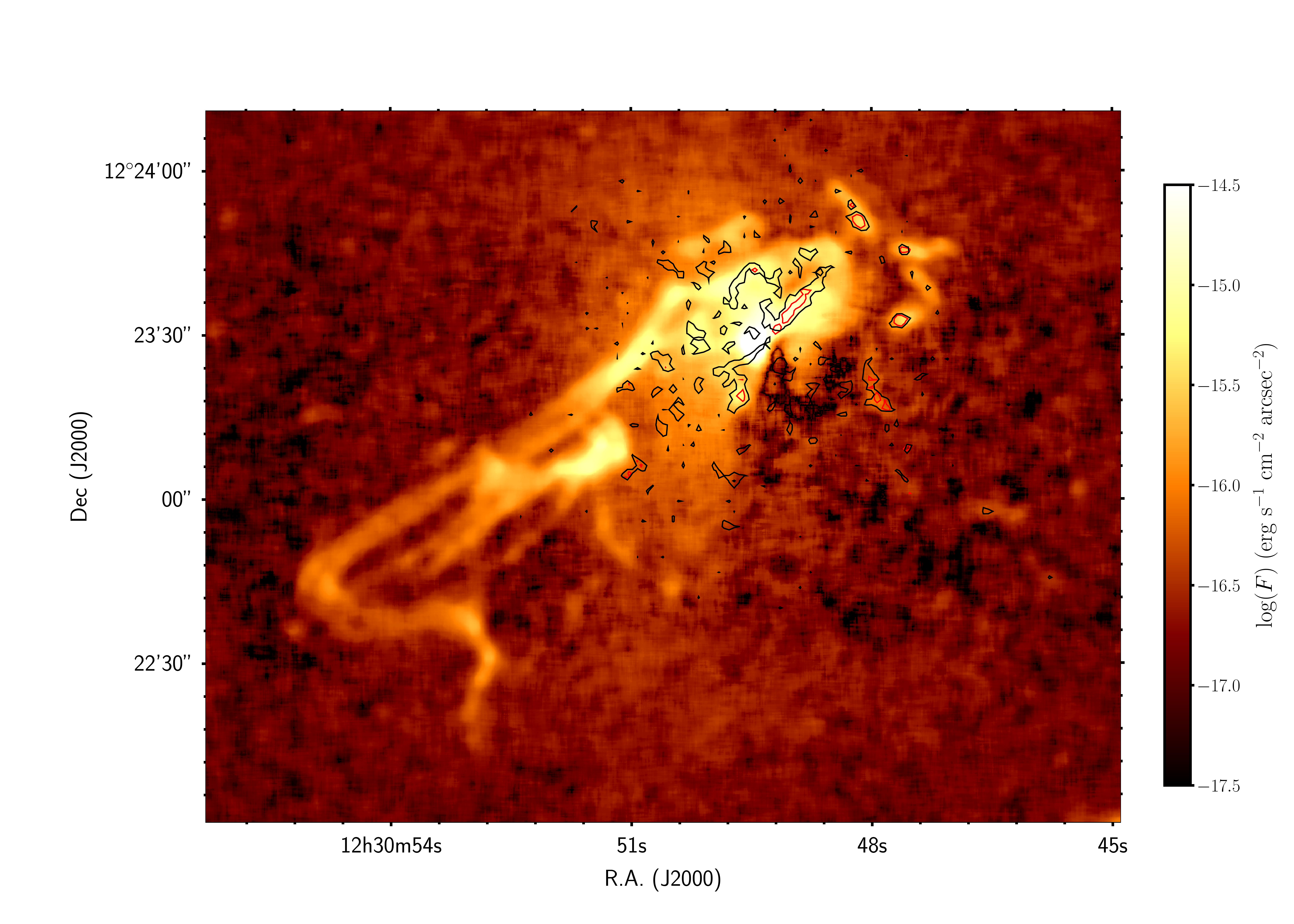}
   \caption{Comparison between the H$\alpha$+[NII] emission (in colour) and the dust distribution (contours) derived from the NGVS image in absorption.
 }
   \label{M87Hadust}%
   \end{figure}

Figure \ref{M87Hadust} shows the comparison between the ionised gas distribution traced by the H$\alpha$+[NII] and the distribution of the
dust component as derived by the NGVS image (see Sect. 3.3). As already noticed by Sparks et al. (1993), there is a quite clear association between 
the north-west filament located just on the northern side of the jet and a dust lane. The extremely deep NGVS image allowed us to detect 
several other dust features previously unknown. Clearly most of them are also associated with prominent ionised gas filaments, as for instance is 
the structure at $\simeq$ 14 arcsec ($\simeq$ 1 kpc) to the north of the nucleus called ``bar structure" in Sparks et al. (2004), although dust and ionised gas are here distributed 
in a different way, the former on a circular loop, the latter along an elongated structure. Dust is also present on the ``patch" at $\simeq$ 10 arcsec ($\simeq$ 800 pc) south to the nucleus, 
and in a bright region at $\simeq$ 10 arcsec ($\simeq$ 800 pc) to the west. There is also a clear association 
in the four ionised gas filaments to the north-west, west, and south-west of the nucleus, at roughly 32-to-24 arcsec (2.6-1.9 kpc).
Because the surface brightness of the galaxy rapidly decreases outward, we are not able to detect dust in absorption at distances larger than $\simeq$ 30-35 arcsec (2.4-2.8 kpc), we thus do not know
whether the long south-eastern filament is associated with any dusty feature.

\begin{figure}
   \centering
   \includegraphics[width=0.5\textwidth]{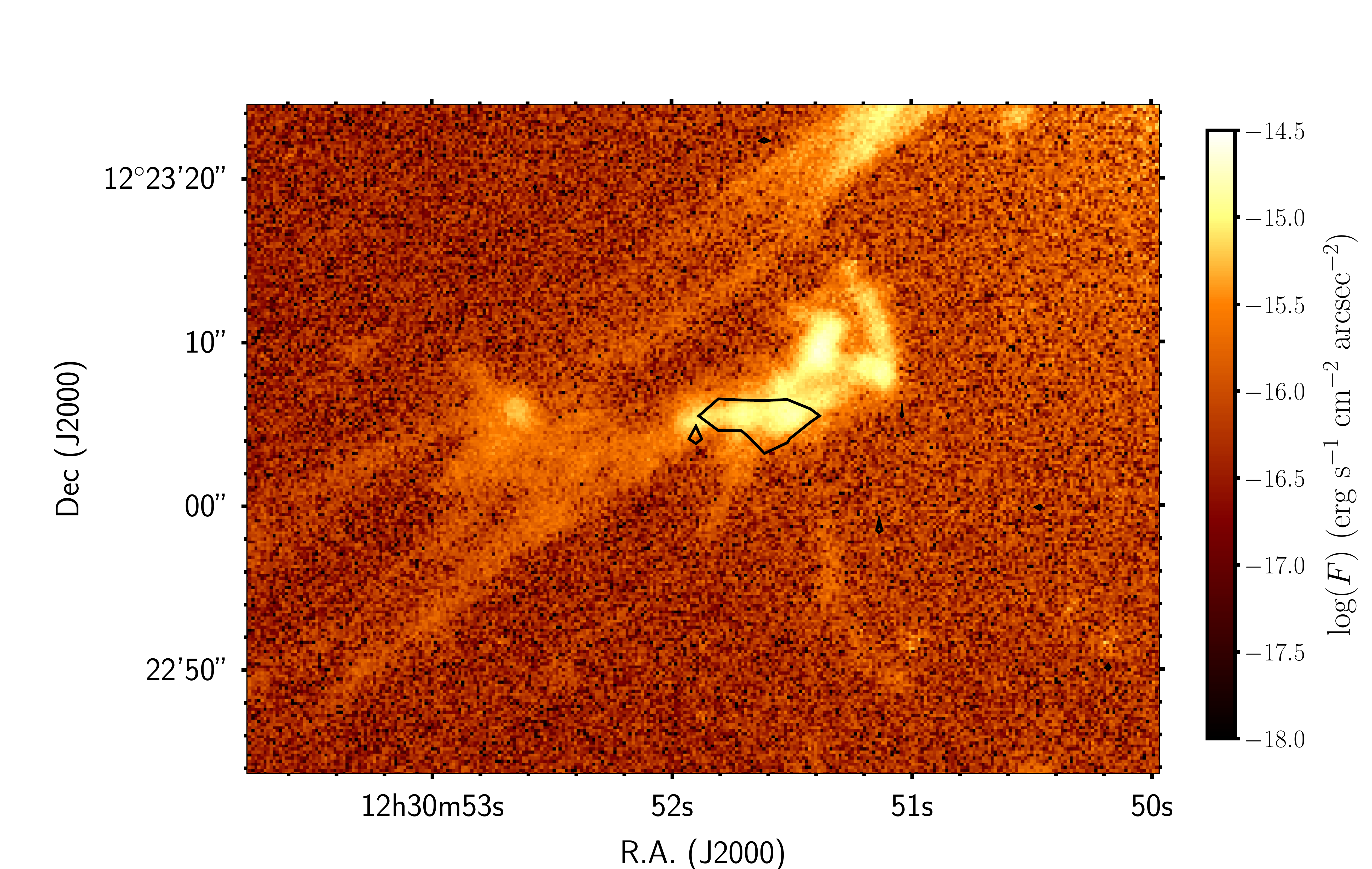}
   \caption{The ALMA $^{12}$CO(2-1) line emission contour at 2 mJy beam$^{-1}$ overplotted on the H$\alpha$+[NII] image. The molecular gas
   has been detected at $\sim$ 40" south-east from the nucleus of M87. The H$\alpha$ filament is located at the south-east edge of the 
   cavity formed by the counter-jet. 
 }
   \label{ALMACO}%
   \end{figure}

We can also compare the distribution of the ionised gas to that of the cold gas component derived from the $^{12}$CO(2-1) line by Simionescu et al. (2018)
using ALMA data and from the [CII] line by Werner et al (2013). Since the gas has been detected only on a very limited region at the edge of the counter-jet in the south-eastern direction, 
we show the comparison on a zoomed region (Fig. \ref{ALMACO}). The cold gas, which has a mass of $M(H_2)$ $\simeq$ 5 $\times$ 10$^5$ M$_{\odot}$, 
is located on a peak of ionised gas emission. We do not see, however, any associated compact source in 
both the broad- and narrow-band images, indicating the lack of any star forming system. 

\begin{figure}
   \centering
   \includegraphics[width=0.5\textwidth]{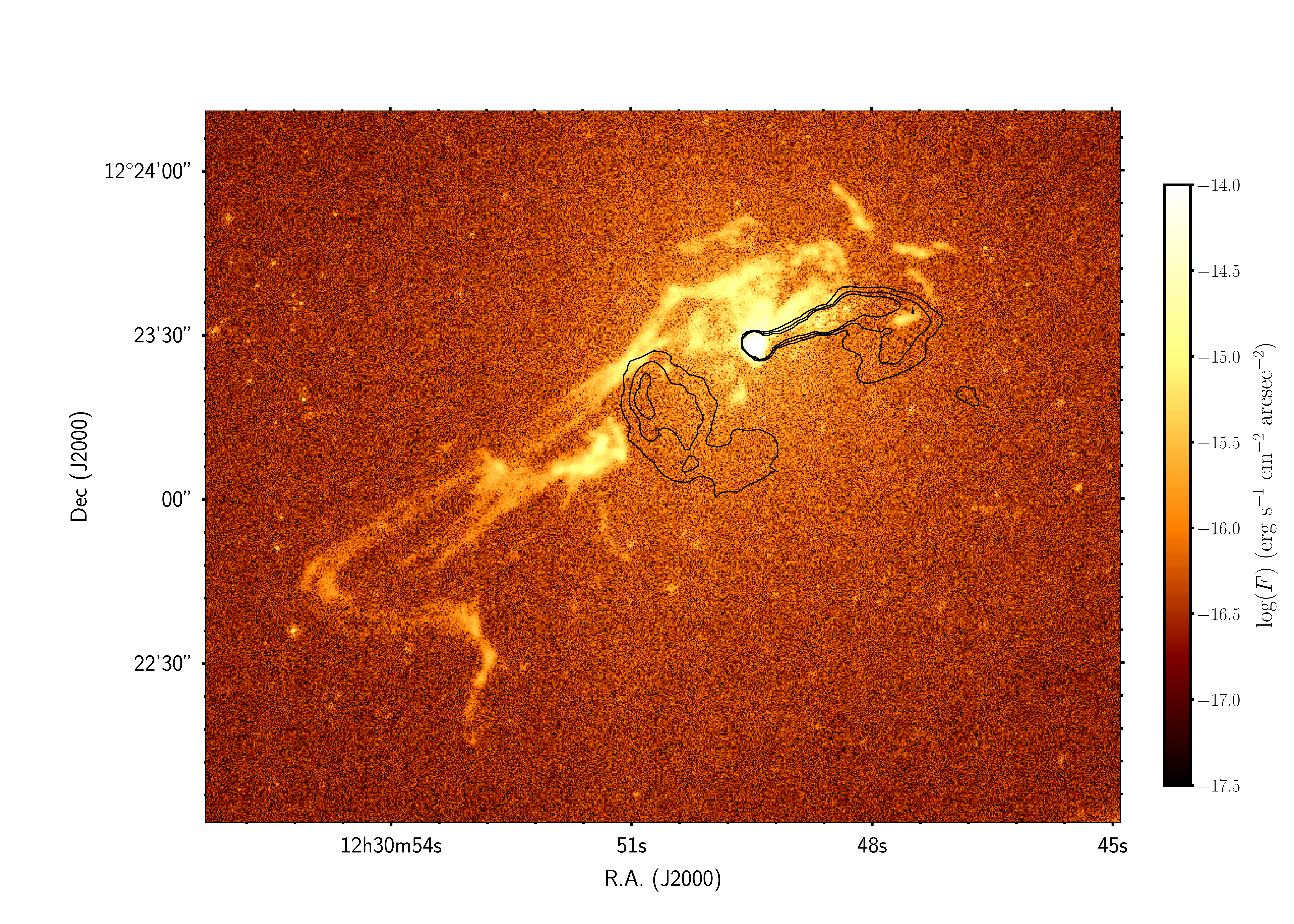}
   \caption{ALMA contours in the radio continuum at 115 GHz of M87 at levels of 1.5, 5, 10 mJy beam$^{-1}$ P
   overplotted on the continuum-subtracted H$\alpha$+[NII] image.
 }
   \label{M87HaALMAcont}%
   \end{figure}

\begin{figure}
   \centering
   \includegraphics[width=0.5\textwidth]{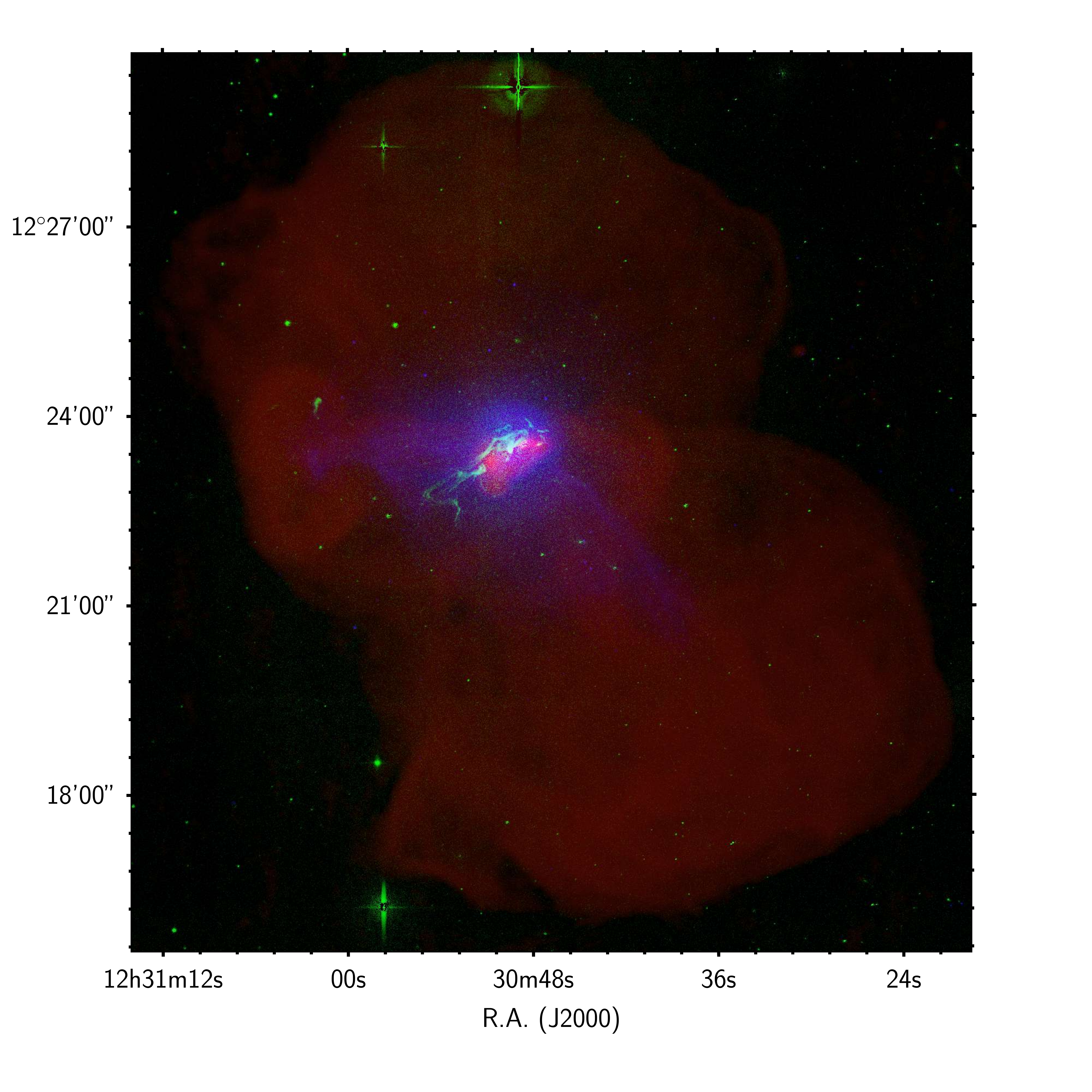}
   \caption{The pseudo-colour image of M87 obtained by combining \textit{Chandra} 1.0-3.5 keV (blue),
   VESTIGE H$\alpha$+[NII] (green),  and the VLA radio continuum at 90 cm (red) frames of the galaxy. 
 }
   \label{multifrequency}%
   \end{figure}

Figure \ref{M87HaALMAcont} compares the ionised gas emission to the radio continuum emission at 115 GHz. As already noticed in previous works, the 
ionised gas emission is located northern to the prominent jet, and avoids the counter-jet lobe in the south-eastern direction.  
As a final image, we show in Fig. \ref{multifrequency} the pseudo-colour image of M87 obtained by combining \textit{Chandra} 1.0-3.5 keV,
H$\alpha$+[NII] and radio continuum at 90 cm frames of the galaxy. This figure allows us to appreciate how the emission at these different wavelengths 
correlates at large scale. Again we see how the H$\alpha$ emission avoids the main radio structures in the inner regions (Sparks et al. 1993), while
the ionised gas plume and filaments at $\simeq$ 15 and 18 kpc east from the nucleus are located within the radio lobe (Gavazzi et al. 2000).

\section{Discussion}

The detailed discussion of the properties of the different gas phases of M87 has been the topic of several publications (e.g. Sparks et al. 1993, 2004, Forman et al. 2007, 2017, 
Churazov et al. 2008, Werner et al. 2010, 2013, Simionescu et al. 2018). We thus refer the reader to these works. Here we just focus on the main novelties provided by our data, mainly due to
their higher sensitivity and angular resolution with respect to those gathered in previous works.

\subsection{The source of ionisation}

The exquisite quality of the MUSE data allowed us to make a 2D BPT diagram along the ionised gas filament and compare it to the prediction of the photoionisation models of Kewley et al. (2001). 
Figure \ref{BPT} (upper panels) clearly shows that, in the case of photoionisation, the [NII]/H$\alpha$ line ratio is expected to increase with increasing metallicity, 
but it never reaches the values observed in the gas filaments of M87. We recall that in Virgo the metallicity of the ICM derived from ASCA is only 0.39-0.55 $Z_{\odot}$
(Matsumoto et al. 2000; Fukazawa et al. 2004). Photoionisation due to young massive stars does not seem to be at play even in the
prominent ionised gas region at the south-eastern edge of the counter-jet, where molecular gas has been detected (Fig. \ref{ALMACO}). Indeed, we do not
observe any compact stellar counterpart in both the UV and optical images, we thus deduce that there is no star formation
associated with this molecular clouds. We recall that Simoniescu et al. (2018), using dynamical arguments, concluded that because of its velocity dispersion of 27 km s$^{-1}$,
the molecular gas is here stable, and it is thus highly unlikely that it can collapse to form new stars. Positive feedback does not seem to be present in this object.
This result is consistent with what found by Johnstone et al. (2007) after comparing the predictions of photoionisation models with 
the optical and infrared line properties of cooling-flow galaxies in the core of clusters. Other sources of ionisation have been proposed, such as the pure collisional heating due 
to cosmic rays accelerated by magnetohydrodynamic waves present along the filament (Ferland et al. 2009, McDonald et al. 2012), dust scattering
(Mattila et al. 2007, Seon \& Witt 2012), thermal conduction from 
the hot phase (Sparks et al. 2004), or ionisation due to the central AGN. The deep optical image shown in Fig. \ref{NGVS} indicates that dust is not ubiquitously distributed along the filament,
contrary to what found in other bright central galaxies (e.g. NGC 4696, Crawford et al. 2005). Ionisation from the AGN might be important in the central 0.5 kpc region
as suggested by the radial gradient of the [NII]/H$\alpha$ line ratio with radius and the relatively high [OIII]/H$\alpha$ ratio (Fig. \ref{BPT}),
while it is probably negligible in the outer filament (McDonald et al. 2012).
Our analysis also suggests that the gas might be ionised by shocks (Fig. \ref{sigma2}).
The distribution of the ionised gas at the edges of the radio jet or 
of the counter-jet might indicate that the shocks can be produced by the energy injected by the AGN in the ISM. The contribution of shocks to the ionisation of 
the prominent gas filaments of M87 seems more important than in other central galaxies in cooling flow clusters given its higher [OIII]/H$\alpha$ line ratio (McDonald et al. 2012).

\subsection{The origin of the filament}

The kinematical properties of the gas (Fig. \ref{vel}) compared with those recently derived from the planetary nebulae (Longobardi et al. 2015,
2018) and the globular clusters (Romanowsky et al. 2012) in the outer
halo, from the stellar kinematics in the inner regions (Emsellem et al. 2014, Arnold et al. 2014), or from the observed substructures in the colour 
distribution of the globular clusters (Powalka et al. 2018), seem 
to suggest that the ionised gas filament is the remnant of a gas-rich galaxy which has been recently accreted by M87, as first proposed by Sparks et al. (1993),
or has crossed the halo loosing most of its gas reservoir after a ram pressure stripping event (Mayer et al. 2006). The rough estimate of the total mass of the ionised gas in the filament 
($M_{filament}$ $\simeq$ 6.9 $\times$ 10$^7$ M$_{\odot}$) is comparable to that of the gas that would have been associated with the observed dusty filaments if this dust had been accreted from an external object
($M_{gas,dust}$ $\simeq$ 2 $\times$ 10$^7$ M$_{\odot}$). This gas mass, however, should be considered as a lower limit to the total gas mass lost by the star-forming galaxy because 
dust might have been destroyed by sputtering in the harsh intertsellar radiation field of M87 (Goudfrooij \& de Jong 1995, Temi et al. 2007).
The kinematical properties of $\sim$ 300
planetary nebulae in the outer halo and a detailed colour analysis of the galaxy led Longobardi et al. (2015, 2018) to conclude 
that M87 has recently ($\lesssim$ 1 Gyr) accreted a massive ($M_{star}$ $\simeq$ 6 $\times$ 10$^9$ M$_{\odot}$) star forming system. 
The comparison between the distribution of $\sim$ 500 globular clusters in the phase diagram with tuned simulations suggests that the galaxy 
has been continuously accreting smaller systems for $\sim$ 1 Gyr (Romanowsky et al. 2012). Furthermore, M87 has also a kinematically distinct component in the inner $\sim$ 1.2 kpc radius 
which might have been formed after a gravitational perturbation (Emsellem et al. 2014). Accretion of satellites is also suggested by the 
low surface brightness fine structures in the deep optical image of M87 (Mihos et al. 2017).
The complex kinematics along the filament rules out a simple scenario where the gas has been lost while the star forming progenitor was spiraling around M87,
as observed in the central galaxy of the cluster A2052 (Balmaverde et al. 2018), and get virialised on short timescales as expected in massive objects. Indeed, the energy
ejected by the AGN can mix, excite, and accelerate the gas which is conserving its angular momentum while infalling within the potential well of the galaxy, 
producing the peculiar velocity field shown in Fig. \ref{vel}.
Complex structures in the velocity field of the ionised gas filaments have been observed in other radio galaxies such as NGC 1275 in Perseus (Gendron-Marsolais et al. 2018),
where star formation is also present (Canning et al. 2014).

An alternative interpretation is that, as in other central cluster galaxies, the ionised gas filament might be produced by the diffuse hot gas halo which, for local instabilities, 
cools out to form magnetically supported narrow filaments.
The velocity map shown in Fig. \ref{vel} reveals generally negative velocities for the gas adjacent to the jet and positive velocities
on the other side. This is in general consistent with the scenario that at least some ionised emitting gas interacts
with jets and the counter-jet. Indeed, the ionised gas filaments of M87 avoids the radio bubbles (Figs. \ref{M87HaALMAcont} and \ref{multifrequency}) as often observed in cooling-flow galaxies (Russell et al. 2017).
Jets can trigger buoyant bubbles filled by radio plasma, rising in the X-ray atmosphere of M87 (Churazov et al. 2001).
Such kind of rising bubbles can have velocities of 60\% - 70\% of local sound speed, 
or 300 - 470 km/s for kT = 1 - 1.8 keV (Million et al. 2010).
The ionised gas filaments at the south-east align with the current jet / counter jet, and might have been formed out of thermal
instabilities in uplift gas from the central region by radio bubbles (e.g., Churazov et al. 2001; Forman et al. 2007).
The ionised gas inside the radio contours in the western direction might thus be at the edges of the radio bubble (projection effects).


\section{Conclusions}

The analysis of deep narrow-band H$\alpha$+[NII] images of the elliptical radio galaxy M87 in the centre of the Virgo cluster confirmed the presence of an ionised gas filament 
extending up to 3 kpc in the north-western and 8 kpc in the south-eastern direction. It also allowed the detection of a low surface brightness plume of ionised gas at 
$\simeq$ 18 kpc to the east of the nucleus, located at the edge of the radio continuum counter-jet lobe and corresponding to a sharpe edge in the 0.5-1.0 keV X-rays emission.
The kinematical properties of this gas filament derived using new MUSE IFU data revealed the presence of a very perturbed velocity field, with differences in 
velocity as high as $\sim$ 800 km s$^{-1}$ on scales $\lesssim$ 1 kpc, probably due to different gas components located along the line of sight. 
The position of the spaxels along several diagnostic diagrams, and the observed relationship between the shock-sensitive [OI]/H$\alpha$ line ratio and the velocity dispersion of the gas,
consistently suggest that the gas is shock-ionised probably by the expanding bubbles of the rotating radio jet. The mass of the ionised gas in the filament, or that of the
cold gas expected to be associated with the dust seen in absorption (both a few 10$^7$ M$_{\odot}$) might suggest that the filament is a remnant of gas coming 
from a galaxy recently captured by M87.
The ionised gas filament, however, could also be gas cooled out from the hot gas halo along magnetically supported narrow filaments.

\begin{acknowledgements}

We thanks the anonymous referee for constructive comments. We thank N. Grosso for his help in gathering the \textit{Chandra} X-ray data of M87.
We are grateful to the whole CFHT team who assisted us in the preparation and in the execution of the observations and in the calibration and data reduction: 
Todd Burdullis, Daniel Devost, Bill Mahoney, Nadine Manset, Andreea Petric, Simon Prunet, Kanoa Withington.
We acknowledge financial support from "Programme National de Cosmologie and Galaxies" (PNCG) funded by CNRS/INSU-IN2P3-INP, CEA and CNES, France,
and from "Projet International de Coop\'eration Scientifique" (PICS) with Canada funded by the CNRS, France.
This research has made use of the NASA/IPAC Extragalactic Database (NED) 
which is operated by the Jet Propulsion Laboratory, California Institute of 
Technology, under contract with the National Aeronautics and Space Administration
and of the GOLDMine database (http://goldmine.mib.infn.it/) (Gavazzi et al. 2003).
This project has received funding from the European Research Council (ERC) under the European Union's Horizon 2020 research and innovation programme (grant agreement No 757535 and
grant agreement No 679627, project name FORNAX).
MB acknowledges the FONDECYT regular grant 1170618.
MS acknowledges support from the NSF grant 1714764.

\end{acknowledgements}

\end{document}